%% file: 00_Article_Merge.tex
\documentclass[preprint]{bioRxiv}

\usepackage{lipsum}
\usepackage{amsmath,amssymb}
\usepackage{mathtools}

\begin{document}

\include{01_Article_MainText}
\include{02_Article_Supplementary}

\end{document}

%% file: 01_Article_MainText.tex
\leadauthor{Sridhar}

\title{Uncovering multiscale structure in the variability of larval zebrafish navigation} 

\shorttitle{Uncovering timescales and variability in complex motor patterns}

\author[1]{Gautam Sridhar}
\author[2]{Massimo Vergassola}
\author[3]{João C. Marques}
\author[3]{Michael B. Orger}
\author[1,2,\Letter]{Antonio Carlos Costa}
\author[1,\Letter]{Claire Wyart}
\affil[1]{Sorbonne University, Paris Brain Institute (ICM), Inserm U1127, CNRS UMR 7225, Paris, France}
\affil[2]{Laboratoire de Physique de l'Ecole normale supérieure, ENS, Université PSL, CNRS, Sorbonne Université, Université de Paris, F-75005 Paris, 
France}
\affil[3]{Champalimaud Research, Champalimaud Centre for the Unknown, Avenida Brasília, Doca de Pedrouços, Lisboa 1400-038, Portugal}
\date{}

\maketitle

\begin{abstract}

Animals chain movements into long-lived motor strategies, exhibiting variability across scales that reflects the interplay between internal states and environmental cues. To reveal structure in such variability, we build Markov models of movement sequences that bridges across time scales and enables a quantitative comparison of behavioral phenotypes among individuals. Applied to larval zebrafish responding to diverse sensory cues, we uncover a hierarchy of long-lived motor strategies, dominated by changes in orientation distinguishing cruising versus wandering strategies. Environmental cues induce preferences along these modes at the population level: while fish cruise in the light, they wander in response to aversive stimuli, or in search for appetitive prey. As our method encodes the behavioral dynamics of each individual fish in the transitions among coarse-grained motor strategies, we use it to uncover a hierarchical structure in the phenotypic variability that reflects exploration-exploitation trade-offs. Across a wide range of sensory cues, a major source of variation among fish is driven by prior and/or immediate exposure to prey that induces exploitation phenotypes. A large degree of variability that is not explained by environmental cues unravels motivational states that override the sensory context to induce contrasting exploration-exploitation phenotypes. Altogether, by extracting the timescales of motor strategies deployed during navigation, our approach exposes structure among individuals and reveals internal states tuned by prior experience.

\end{abstract}


\begin{corrauthor}
Antonio Carlos Costa, antonio.costa@icm-institute.org; Claire Wyart, claire.wyart@icm-institute.org
\end{corrauthor}



\section*{Introduction}\label{s:introduction}

Animal behavior emerges from the interplay between external environmental cues and internal states incarnated in complex biological processes such as interoception, neuromodulation and hormonal regulation (\cite{bargmann2012beyond,marder2012neuromodulation,kanwal2021internal}), leading to the emergence of structured behavioral variability across individuals in a population (\cite{deBivort2022precise}). The common \emph{reductionist approach} of focusing on population averages in biology can lead to errors due to averaging (see an early example in \cite{golowasch2002failure} and a large scope review in \cite{marder2023individual}). Applied to behavior, averaging selected kinematic parameters forbids to explore and to quantify the whole spatiotemporal structure to the variability among individuals (\cite{jacobs2023larval}). One major challenge lies in quantitatively assessing the time scales at which external and internal influences shape behavioral variability. This point is the core motivation for our study: we develop an approach to quantify the hierarchy of timescales in behavior, revealing structure to behavioral variability in large populations of animals exposed to various environmental cues.

\paragraph{}
Most attempts to quantify behavior focus on particular spatiotemporal scales of interest (\cite{schwarz2015_changes, marques2018structure, helms2019modelling}). Recent analysis of short timescale behaviors at high resolution has for example shed light on the neuronal circuits underlying sensorimotor integration (\cite{meng2024tonically,kano2023awc,darmohray2019spatial,bohm2016csf,knafo2017mechanosensory,carbo2023mesencephalic}). In contrast, the analysis of long timescale behaviors at low spatiotemporal resolution uncovered changing internal states (\cite{ben2009molecular, fujiwara2002regulation,flavell2013serotonin, marques2020internal}). However, external and internal influences typically manifest themselves in behavior across several timescales. It is therefore essential to develop holistic approaches that can disentangle the multi-dimensional structure of behavior across scales, from posture movements to search strategies. 

\paragraph{}
Recent advances in machine vision have enabled an unprecedented lens onto the multiple scales of behavior across species (\cite{brown2018ethology,berman2018measuring,datta2019computational,pereira2020quantifying}), allowing for high resolution posture measurements in large environments while spanning several timescales -- from milliseconds to hours. We aim to bridge across timescales in behavior by constructing predictive models that can capture statistics of long timescale motor strategies from the integration of fine-scale movements. Markov models,in principle, offer that possibility: they not only have the potential to be accurate predictive machines, but they also offer a lens into the long-lived properties of behavior through the eigenvalues and eigenvectors of the transition matrix. However, building accurate Markov models requires predictive representations of behavior. Numerous studies in genetic model organisms relied on unsupervised approaches to identify a small number of stereotyped movements, such as "bout types" (\cite{marques2018structure, mearns2020deconstructing, johnson2020probabilistic}) or "syllables" (\cite{wiltschko2015mapping, wiltschko2020revealing, weinreb2023keypoint}). Such behavioral categorization might however unavoidably erase fine-scale information, making the prediction of long timescale sequences challenging. Accordingly, a Markov model built from bout types is unable to uncover the frequency with which larval zebrafish engage in long sequences of bouts (\cite{reddy2022lexical}), pointing to non-Markovianity. This is not an isolated observation: non-Markovianity appears in behavioral data across species (\cite{schwarz2015_changes,berman2016predictability}) and cannot be decoupled from the behavioral representation.


\paragraph{}
We thus carefully conceive our analysis to retain fine scale kinematics and history dependence. As recent work has shown (\cite{costa2023markovian}), taking these aspects into account allows the bridging of fine scale posture movements to longer lived motor strategies within the same Markov model. Using this multiscale approach to behavior, we investigate how sensory information and internal states drive behavior across timescales. We take advantage of the power of larval zebrafish, which are small enough to record multiple individuals for a long duration (\cite{orger2017zebrafish}). We apply our approach on previously collected datasets in which 6-7 days post-fertilization (dpf) larval zebrafish are exposed to a large variety of stimuli (\cite{marques2018structure,reddy2022lexical}). To design a model that accurately capture the long-lived properties of larval zebrafish behavior, we study the dynamical evolution of behavior across maximally-predictive bout sequences. We obtain Markov models that are predictive of the behavioral dynamics of each fish across timescales. 

\paragraph{}
Using parsimonious yet predictive descriptions of an individual animal's behavior across this hierarchy of timescales enables us to develop a novel approach to dissect individual variability and reveal structure in a large population of animals, disentangling how external sensory cues and persistent hidden states drive behavior. Using models encoding  individual behaviors, we reconstruct a phenotypic space that captures overall tendencies for different long-lived motor strategies across fish. We find that the structure of this phenotypic space is only partially determined by sensory contexts, pointing to hidden internal variables. Accordingly, fish with different phenotypic biases exhibit differential sensorimotor transformations across multiple timescales. Surprisingly, prior or current exposure to prey has the most profound impact on behavioral variability. Through simulations, we discover that the persistent phenotypic preferences are well suited for either pursuing and capturing prey, engaging in local searches or performing large distance dispersal, reflecting fundamental exploration-exploitation trade-offs.

\section*{Results}

\subsection*{Uncovering the multiple scales of larval zebrafish behavior}

\begin{figure*}[ht]
\centering
\includegraphics[width = \textwidth]{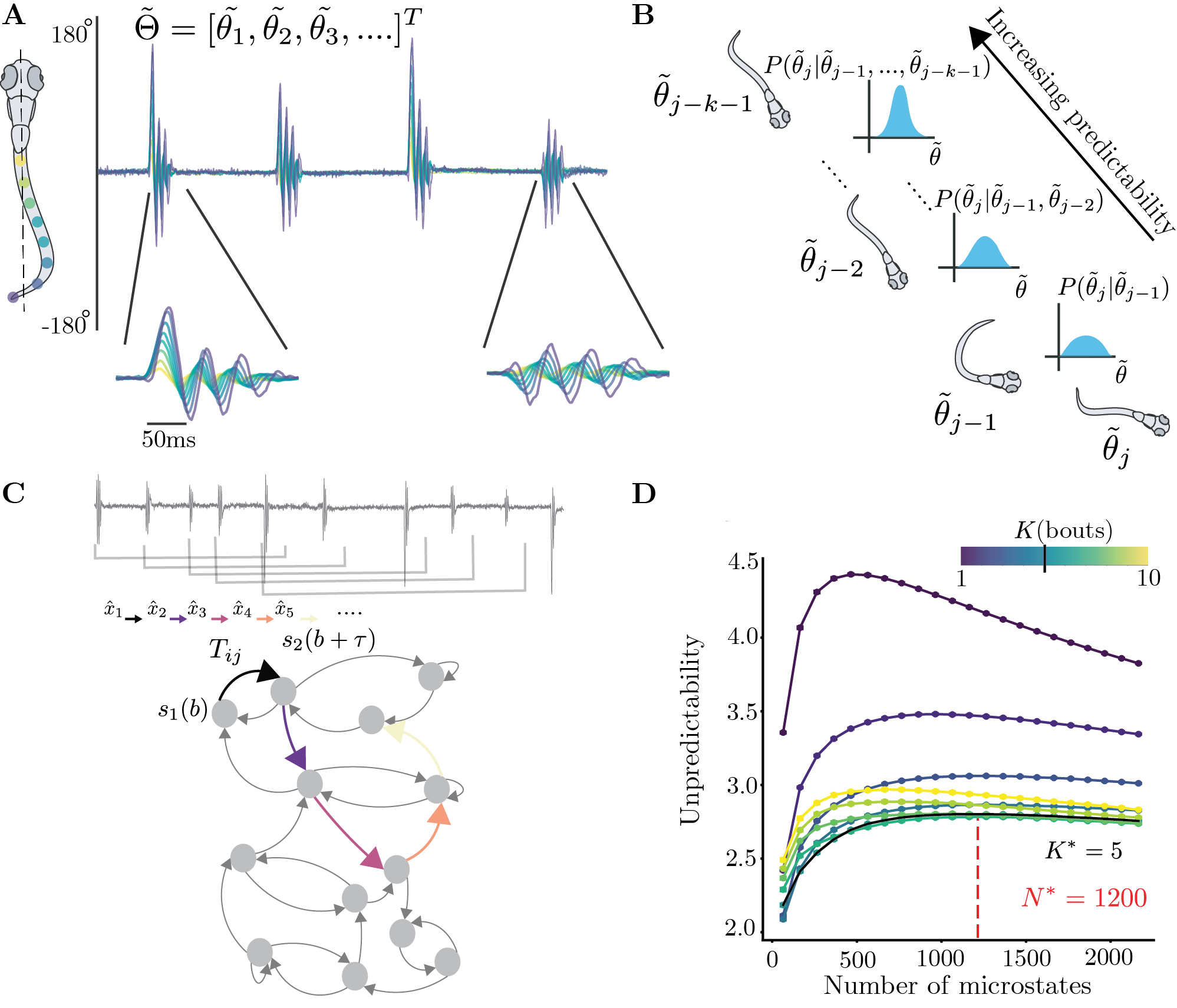}
\caption{\textbf{Building a maximally predictive state space for larval zebrafish} \textbf{(A)} Typical example of larval zebrafish locomotion bouts, measured as the angle made by points on the tail with respect to the midline. (\textbf{below}) Examples of a turn and forward bout. \textbf{(B)} In order to recover the underlying dynamics in the bout space, we concatenate bouts from the past behavior of the fish. With more information about the past bouts, the predictability of the future bout is expected to increase. \textbf{(C)} The fish's locomotion is then characterized by a sequence of overlapping windows of $K$ bouts $\hat{x}_t = \{\tilde{\theta}_t,\ldots,\tilde{\theta}_{t+K-1}\}$ (here $K=5\,\text{bouts}$). The dynamics of this sequence can be encoded in a Markov chain which we build by discretizing the space of bout sequences into $N$ microstates, obtaining $s_i\,i\in\{1,\ldots,N\}$ (see Methods). The unpredictability of the sequence can then be quantified by the short-time entropy rate of the Markov chain. \textbf{(D)} The entropy rates (in nats/bout) of Markov chains for the data from \cite{marques2018structure}. We sample 7500 bouts from each sensory context equally and estimate the entropy rate of ensemble markov chains at various $K,N$. Note the decreasing entropy rate with respect $K$, minimizing at $K^*=5$. To maximize the amount of fine-scale dynamics captured by our Markov chain, we pick the maximum number of partitions $N$ according to the entropy rate (shown here at $N^*=1200$ for data from \cite{marques2018structure}).}
\label{fig:1}
\end{figure*}

Larval zebrafish move via sub-second tail oscillations, referred to as ``bouts'', which are separated by periods of rest lasting typically $\approx 0.5 \,\text{s}$ (Fig.\,\ref{fig:1}A, see Methods). While the tail beats on timescales of $O(10^{-2}\,\text{s})$ during a bout, stereotyped sequences of bouts last several seconds to minutes (\cite{reddy2022lexical}). We take advantage of large published datasets (\cite{marques2018structure}) in which $463$ freely-swimming larval zebrafish were recorded from 30 minutes to 3 hours under a variety of sensory contexts. In what follows, sensory context refers to the immediate stimulus experienced by the animal (e.g., navigation in the light, in the dark, looming stimulus, prey capture, etc.), but also the geometry of the arena. Certain groups of fish also experience a different prior context, which involves being raised with live food from 3 days post fertilization (dpf) (see Table.\,\ref{Suppl:Table}). We utilize the posture of the tail during a bout until for a duration of up to $250\,\text{ms} = 175\,\text{ frames}$ after bout initiation, so that each bout is encoded as a $8\,\text{tail angles} \times 175\,\text{ frames}$ dimensional object. To work with a concise low dimensional representation of bouts and filter out noisy bout features, we perform principal component analysis (PCA) on all bouts, and keep 20 dimensions which explain over $95\%$ of the variance across bouts (Fig.\,\ref{Suppl:1}, see Methods). 

\paragraph{}
To reveal the long-lived properties of larval zebrafish navigation, we adapt the approach of maximally predictive Markov models introduced in \cite{costa2023maximally}. We resolve the history-dependence of the dynamics by including past bouts into an expanded representation (Fig.\,\ref{fig:1}B) of the behavioral dynamics. We sample 7,500 bouts from each of the 14 sensory contexts and search for a maximally-predictive representation of the \emph{ensemble} dynamics.  
We built sequences of bouts (Fig.\,\ref{fig:1}C), increasing the sequence length to maximize the predictability of a Markov chain built from clustering the bout sequences into $N$ microstates (see Methods). We assess predictability by estimating the entropy rate of the resulting Markov chain, which reflects the variability in future states given current states as a function of the number of bouts in a sequence, $K$, and the number of microstates $N$ (Fig.\,\ref{fig:1}D). Past bouts help narrow down our predictions of the future, and a large enough $N$ allows us to retain fine-scale information about each bout's kinematics. For the dataset from \cite{marques2018structure}, we set $K^*=5 $ bouts to minimize the entropy rate of the Markov chain, while simultaneously maximizing information content with $N^*=1200$ microstates (see Methods). Given this choice of $K^*$ and $N^*$, we then obtain an ensemble transition matrix $T_\text{ensemble}(\tau)$ that captures the overall behavior of all fish across sensory contexts and can therefore reveal common long-lived structures in the behavioral dynamics. 

\begin{figure*}[ht]
\centering
\includegraphics[width = 1.\textwidth]{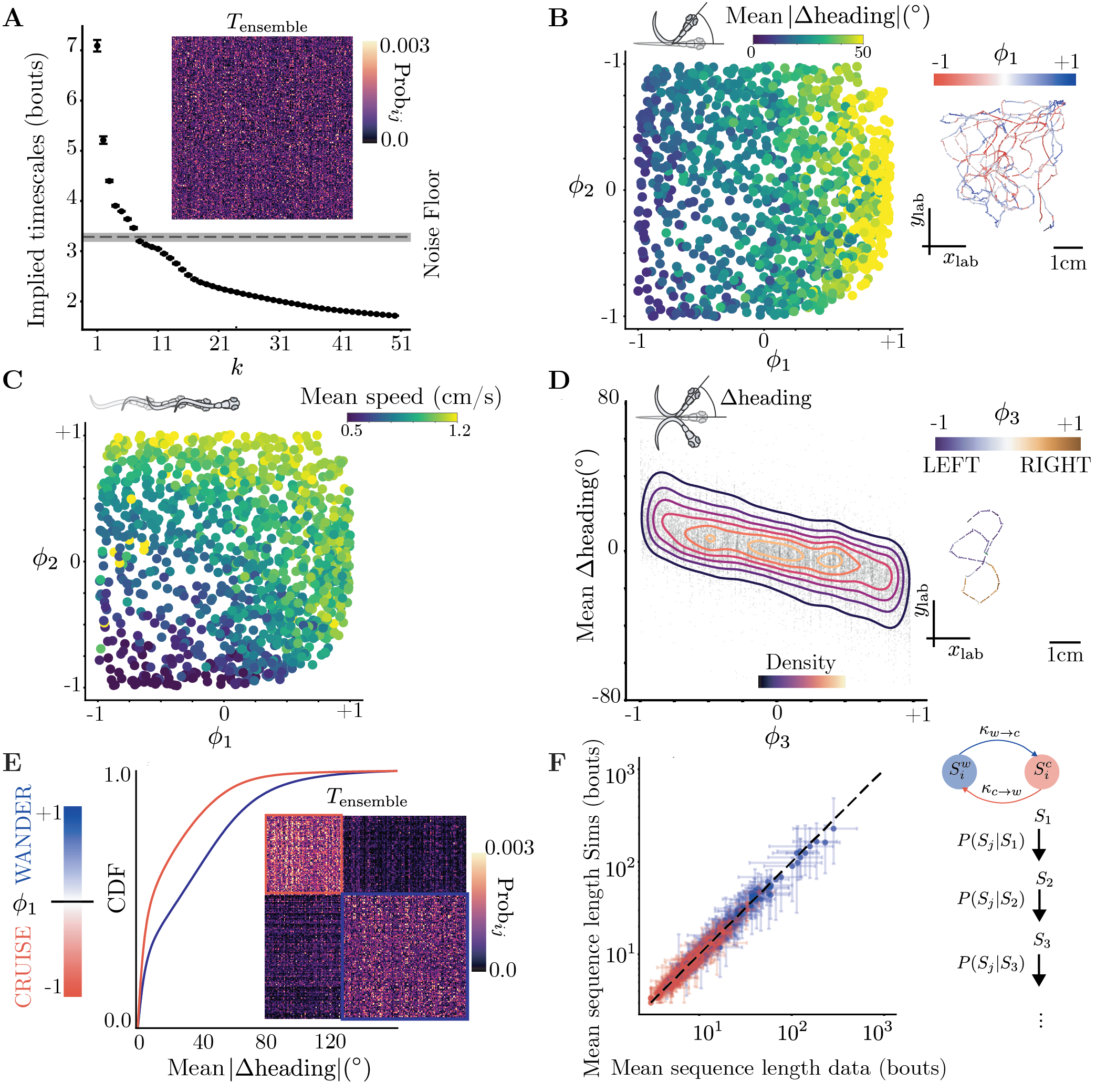}
\caption{\textbf{Navigation is driven by 3 long lived modes in a hierarchy of timescales, prioritizing rate of change of heading, speed and egocentric direction bias} \textbf{(A)} Implied timescales of the $k$-th mode of $T_{\text{ensemble}}$, estimated as $t_\text{imp} = -\frac{\tau^*}{\log \lambda_k}$, where $\lambda_k$ are the eigenvalues of $T_{\text{ensemble}}$ and $\tau^*=3\,\text{bouts}$ (see Methods). $T_{\text{ensemble}}$ is built by sampling 7500 bouts from each sensory context in \cite{marques2018structure} over a 100 seeds and then averaging. We show $T_\text{ensemble}$ as an inset. Error bars represent bootstrapped 95\% confidence intervals of the implied timescales over these 100 seeds. \textbf{(B)} Microstates organized along $\phi_1-\phi_2$, color-coded by the mean absolute change in heading. The longest lived mode $\phi_1$ seems to correlate to absolute change in heading across bout sequences. \textbf{(inset)} An example trajectory of 500 bouts in the lab space color coded by $\phi_1$. \textbf{(C)} Microstates organized along $\phi_1-\phi_2$, color-coded by the mean speed. The second longest lived mode $\phi_2$ correlates the mean speed across bout sequences. \textbf{(D)} The third mode $\phi_3$ encodes an egocentric preference for left or right directions. \textbf{(inset)} An example trajectory of 40 bouts color coded by $\phi_3$. \textbf{(E)} Partitioning the state space along $\phi_1$ reveals \textit{cruising} and \textit{wandering} metastable motor strategies with either low or high changes in heading. We plot the cumulative distribution function (CDF) of the absolute value of the mean change in heading direction for the cruising (red) and wandering (blue) strategies. In the inset, we organize $T_{\text{ensemble}}$ organized according to the coarse-graining, revealing a block diagonal structure. \textbf{(F)} Mean average sequence length of each fish in the data vs simulations. Fish generate highly variable sequence lengths in each strategy $\{S^w,S^c\}$, from a few bouts to a few hundreds. Our Markov model built from coarse-grained cruising-wandering states for each fish accurately predicts the mean sequence length in the metastable motor strategies.}
\label{fig:2}
\end{figure*}

\paragraph{}
The non-trivial eigenvalues $\lambda_k$ of the ensemble transition matrix $T_\text{ensemble}$ and the respective eigenvectors $\phi_k$ provide a lens onto the organization of behavior across timescales\footnote{Prior to estimating the eigenvalues and eigenvectors of the transition matrix, we enforce its reversibility so that $T_\text{ensemble}$ refers to as a ``reversibilized'' transition matrix, whose eigenvectors optimally distinguish metastable states (\cite{Dellnitz1999,froyland2005statistically,froyland2009,costa2023maximally}).} (\cite{costa2023markovian}). To isolate the long-lived modes from the faster bout-level dynamics, we chose the transition time as $\tau^*=3\,\text{ bouts}$ (Fig.\,\ref{Suppl:2}A), obtaining long-lived modes that are well separated from the bulk spectrum (Fig.\,\ref{fig:2}A). As the fish moves, each sequence of bouts corresponds to a value along $\phi_k$. The time evolution of $\phi_k$ therefore offers a parsimonious description of the transitions among coarse-grained, long-lived behaviors. 
\paragraph{}
We interpret the 3 longest-lived modes by comparing them with common kinematic variables of speed and changes in orientation. The longest-lived mode $\phi_1$ roughly correlates with absolute changes in heading direction (Fig.\,\ref{fig:2}B), the second mode $\phi_2$ correlates with reorientation and speed (Fig.\,\ref{fig:2}C), and the third mode $\phi_3$ with an egocentric direction bias (Fig.\,\ref{fig:2}D) that persists across bouts as previously observed (\cite{dunn2016brain}). Note that specifying a single $\phi_k$ is not sufficient to determine the bout kinematics and the mapping of $\phi_1,\phi_2$ onto specific kinematic variables is non-linear. For example, at intermediate values of $\phi_2$, higher values of $\phi_1$ may correspond to increased speeds (Fig.\,\ref{fig:2}C). The long-lived modes of larval zebrafish navigation are therefore complex functions of simple kinematic parameters, reflecting joint modes of control on different timescales.

\paragraph{}
Proceeding from the longest timescales down, we compress the representation of the long-lived dynamics by sequentially identifying $q$ motor strategies using an increasing number of faster-decaying eigenvectors of the ``ensemble'' Markov model across all fish (see Methods). At the longest timescales, we split along $\phi_1$ to identify dynamically coherent behaviors (\cite{costa2023maximally}), uncovering 2 novel motor strategies, \emph{"cruising"} and \emph{"wandering"}, corresponding respectively to a low and high rate of reorientation (Fig.\, \ref{fig:2}E). The organization of $T_\text{ensemble}$ according to the cruising-wandering categorization reveals a block-diagonal structure indicating the metastability of these strategies(Fig.\,\ref{fig:2}E inset). For $q=4$ motor strategies, our method uncovers \textit{slow} and \textit{fast} variations of \textit{cruising} and \textit{wandering}. For $q=7$, we obtain left/right variations of slow wandering, fast wandering and fast cruising as well as slow cruising without a left-right bias (Fig.\,\ref{Suppl:2} D,E,F). 

\paragraph{}
These metastable states are important motor strategies deployed by the fish across sensory contexts: even though these strategies are obtained from an ``ensemble'' model, a $q$-state Markov model built for single fish offers a good generative model of its behavior (see Fig.\,\ref{fig:2}F,\ref{Suppl:3}A,B). For a coarse-graining into $q$ motor strategies, we simulate the coarse-grained behavior of each fish using its individual transition matrix $T_q^f$ (for cruising-wandering a $q=2$-state Markov Chain, see Methods). Simulated bout sequences of motor strategies from such transition matrices $T_q^f$ accurately predicted the characteristic sequence length (in bouts) of each strategy for individual fish (for $q=2$, see Fig.\, \ref{fig:2}F and for $q=4,7$ see Fig.\, \ref{Suppl:3}A,B). The mean sequence lengths in individual fish vary from a few bouts to hundreds of bouts (Fig.\,\ref{fig:2}F). Such wide range of variation in individual fish is also reflected in the distribution of times spent in either ``cruising'' or ``wandering'' across all fish: although cruising persisted $4.04\,(4.02, 4.07)\,\text{bouts}$ and  wandering $4.12\,(4.09, 4.16)\,\text{bouts}$ (mean with 95\% confidence intervals), these distributions are heavy-tailed (Fig.\,\ref{Suppl:2}G1). There is not one characteristic bout sequence length common across fish and sensory contexts for these motor strategies. Similarly, the bout sequence length of finer scale strategies also varies widely (Fig.\,\ref{Suppl:2}G2,G3). Our analysis yields a hierarchy of Markov models built from an increasing number of coarse-grained states $q$ that are \emph{predictive} on increasingly faster timescales (Fig.\,\ref{Suppl:3}A,B). 

\paragraph{}
To verify that our approach is stable across labs and tracking algorithms, we apply it to a smaller dataset of 6-7 dpf fish exposed to chemical gradients (\cite{reddy2022lexical}) and tracked with a different software (\cite{mirat2013zebrazoom}). Our analysis yields similar modes and motor strategies of long-lived behavior (Fig.\,\ref{Suppl:4}, see Methods) indicating that larval zebrafish navigation is organized along a hierarchy of timescales prioritizing rate of reorientation and instantaneous speed. 

\subsection*{Dissecting the role of motor strategies across sensory contexts}

We hypothesize that the large variability among fish could be explained by the usage of these motor strategies in different sensory contexts. As a first assessment, we estimate the probability of visiting different microstates along $\phi_1-\phi_2$ (Fig.\,\ref{fig:3}A1-A4, see Methods). While there is variability within each sensory context, fish exhibit preferences for particular motor strategies depending on the sensory context. Fish freely exploring an arena in the light ($5\times 5\,\text{cm}^2$ arena, $n=10$ fish) mostly perform \textit{fast cruising}: cruising lasts for $5.84\,(3.27, 8.78)\,\text{s}$ and wandering for $1.18\,(1.00, 1.93)\,\text{s}$ (Fig.\,\ref{fig:3}A1, median values with 95\,\% confidence intervals). When the geometry of the arena restricts behavior along one axis ($1\times 5\,\text{cm}^2$, $n=12$), fish still deploy fast cruising for $2.21\,(1.97, 2.33)\,\text{s}$  but also engage in \textit{wandering} for $1.66\,(1.35, 2.31)\,\text{s}$ (Fig.\,\ref{fig:3}A2), due to the arena geometry forcing reorientation at the corners (Fig.\,\ref{Suppl:6}B). In contrast, in response to aversive stimuli, fish mostly performed \textit{wandering} behaviors: fish in the dark ($n=37$ fish, $\approx$ 30 minutes, $5\times 5\,\text{cm}^2$ arena) display a preference for \textit{fast wandering} (lasting $8.37\,(6.84,11.90)\,\text{s}$) over cruising (lasting $ 3.95\,(2.90,5.82)\,\text{s}$) (Fig.\,\ref{fig:3}A3). Similarly, when exposed to an aversive acidic pH gradient localized in space (\cite{reddy2022lexical}), the avoidance response consists of wandering behaviors (lasting $2.43\,(2.15, 2.84)\,\text{s} $) over cruising (lasting $1.99\,(1.69,2.33)\,\text{s}$) (Fig.\,\ref{Suppl:4}E,F).

\paragraph{}
Next, we investigated the motor strategies deployed by larval zebrafish during prey capture. Detailed studies on larval zebrafish hunting behavior have revealed stereotyped sequences comprising of eye convergence to promote binocular vision of prey (\cite{bianco2011prey, mearns2020deconstructing}) followed by a combination of J-turn, Approach Swim and Capture Swim bout types to successfully track and eat prey (\cite{mcelligott2005prey,marques2020internal, mearns2020deconstructing, johnson2020probabilistic}).
In hunting assays, fish noticeably change their navigation: both during prey capture that is estimated by eye convergence but unexpectedly also in between. To capture preys, larval zebrafish hunting paramecia ($2.5\times 2.5\,\text{cm}^2$ arena, $n=65$ fish) opted for slow cruising (Fig.\,\ref{fig:3}A4) that, as expected, primarily comprises J-turns and Approach Swims (Fig.\,\ref{Suppl:2}E). In the inter-hunt exploratory periods, previous analysis relying on bout types suggested similarity with free swimming behavior (Routine Turns and Slow 1/Slow 2 bout types (\cite{marques2018structure,marques2020internal,mearns2020deconstructing}) or simply exploratory bout types (\cite{johnson2020probabilistic}). We uncover that the inter-hunt exploratory navigation strikingly differs from freely swimming associated with fast cruising in the light and is instead dominated by slow wandering (Fig.\,\ref{fig:3}A4) (wandering lasts $4.13\,(3.11, 4.78)\,\text{s}$ and cruising $1.14\,(1.03, 1.28)\,\text{s}$). Our approach indicates that the wandering motor strategy can be deployed for searching in various contexts: either when exposed to aversive cues (darkness: Fig.\,\ref{fig:3}A3, acidic pH: Fig.\,\ref{Suppl:4}) or to appetite cues (prey capture: Fig.\,\ref{fig:3}A4). 

\paragraph{}
To further dissect what fish can achieve by deploying different motor strategies, we leverage the predictive power of our model to generate synthetic lab space trajectories and assess how distinct motor strategies lead to a differential exploration of space (Fig.\,\ref{fig:3}B). We restrict the dynamics to a given metastable strategy and generate artificial 1000 bout-long sequences. We then sample velocity vectors corresponding to these sequences to generate lab space trajectories (see Methods). Given the simulated trajectories for each motor strategy, we examine two different tasks: one in which the fish has a nearby target within its field of view with the eyes converged, and one in which the fish is broadly searching for resources uniformly scattered on different spatial scales. As expected, slow cruising is the most efficient strategy for pursuing and catching prey with eyes converged (Fig.\,\ref{fig:3}B1). For undirected searches (Fig.\,\ref{fig:3}B2), slow and fast wandering strategies are most efficient on mesoscopic scales, with slow wandering being most efficient up to $\approx 10$ body lengths and fast wandering becoming most efficient between $\approx 20$ and $\approx 50$ body lengths. For resources that are scattered on long distances (beyond $\approx 50$ body lengths), fast cruising becomes the most efficient strategy (Fig.\,\ref{fig:3}B2). Our simulation results thus confirm the benefits of distinct motor strategies: fish freely exploring in the light and never exposed to prey mostly engage in fast cruising as a form of long distance dispersal or exploration, while fish pursuing prey engage in slow cruising, and perform slow and fast wandering as effective search strategies or exploitation, either to search for prey, or to escape aversive environments towards safe areas.

\subsection*{Classifying fish using their behavioral dynamics hints at structure beyond the sensory context}

\begin{figure*}[ht]
\centering
\includegraphics[width = 1.\textwidth]{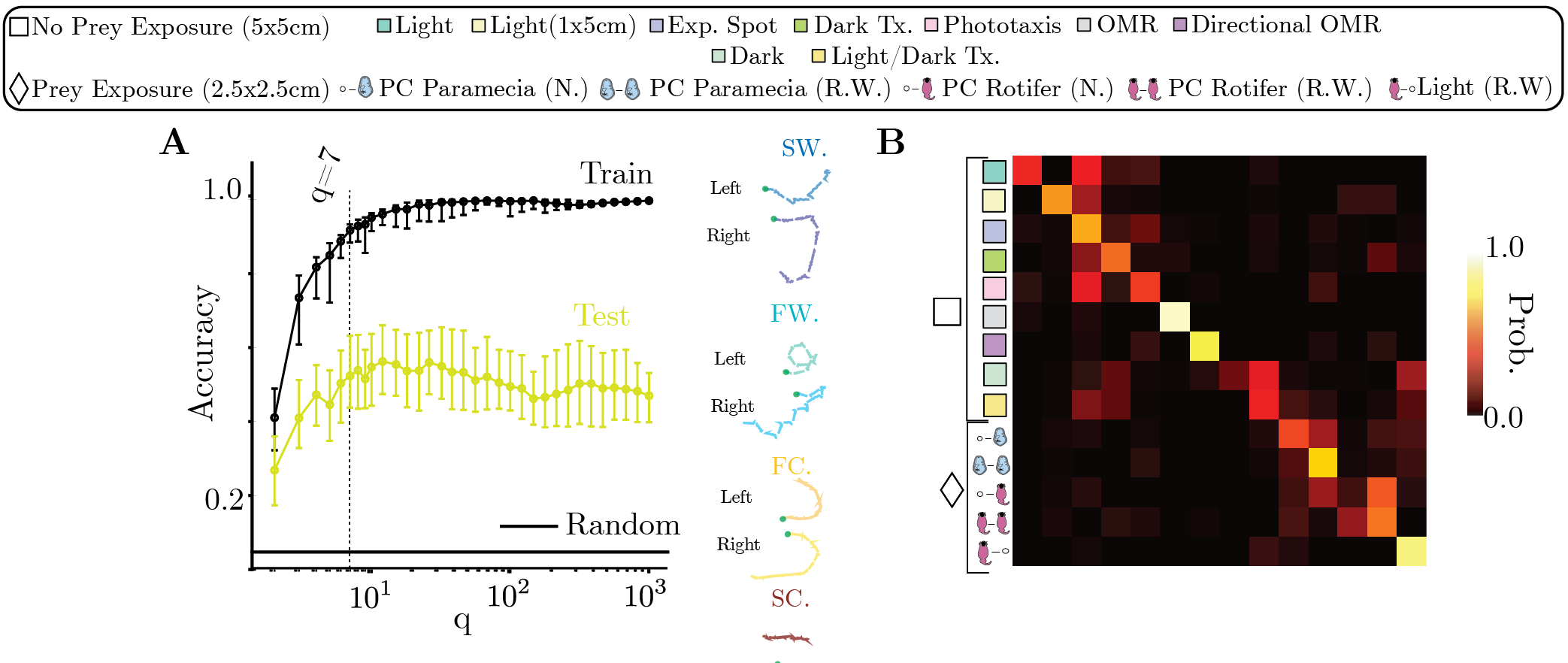}
\caption{\textbf{Sensory contexts are insufficient to explain inter-fish behavioral variability.} We encode the behavior of each fish in transition matrices $T_q^f$ built with an increasing number an coarse-grained states $q$, and use regularized logistic regression to classify each fish into their respective sensory contexts. 
\textbf{(A)} Classification results: accuracy on the Train (black) and Test (yellow) sets as a function of the number of coarse-grained states $q$. We find that the Train accuracy (black) grows continuously as a function of $q$ and quickly reaches $1.0$. The test accuracy (yellow) is much higher than expected from a random label assignment,
but is far from perfect, reaching about $0.5$. It starts decaying past $q\approx 20\,\text{states}$, but with minimal improvements from $q=7\,\text{states}$. This means that while the transition matrix of each fish becomes increasingly unique, allowing for the training accuracy to be nearly perfect, the model quickly fails to generalize to unseen test data: beyond $q\gtrsim 7\,\text{states}$ the classifier starts to overfit. At $q=7$ we find Left-Right variations of slow-fast cruising-wandering motor strategies: see example trajectories on the right.
\textbf{(B)} Confusion matrix of the classifier at $q=7$: each row measures the proportion of individuals from each context that are classified into every condition. A strong diagonal component indicates that many fish are correctly classified into their respective sensory context, while off-diagonal components also indicate that some fish may get misclassified into a different sensory context.
}
\label{fig:4}
\end{figure*}

To delineate the effect of the sensory context on the behavior, we first evaluate whether the sensory context sufficiently explains the behavior of every fish. But to do this unambiguously, we need to assess the adequate coarse-graining scale to compare fish. Intuitively, using too coarse a representation might mix together fish that nonetheless use faster timescale behaviors differently, while using too fine a representation will render each fish unique. We thus classify each fish into their respective sensory contexts by building transition matrices for each fish $T_q^f$ with an increasing number of states $q$, and perform a regularized logistic regression with an 80\%-20\% train-test split (Fig.\,\ref{fig:4}A) (see Methods). The logistic regression shows minimal improvement in accuracy beyond the coarse-graining scale of $q=7$ metastable strategies that split the fast and slow variations of the \textit{cruising-wandering} strategies into egocentric direction preferences (Fig.\,\ref{fig:4}A)(Fig.\,\ref{Suppl:2}C). This result suggests that beyond modulations of reorientation (\emph{cruising-wandering}) and speed (\emph{slow-fast}), persistent \emph{left-right} asymmetry at the experimental timescale also plays a role in distinguishing individual fish from each other across sensory contexts. While the classifier performs better than random, it is an imperfect predictor of sensory context for a given fish as it only reaches a test accuracy $\approx 50\%$ (Fig.\,\ref{fig:4}A). The resulting confusion matrix (Fig.\,\ref{fig:4}B), which measures the probability of a fish to be assigned to a given sensory context, has a strong diagonal component, indicating that most fish can be accurately classified -- especially for optomotor assays that are well-known to reliably drive sensorimotor behavior (\cite{severi2014neural,severi2018investigation}) (center of confusion matrix, Fig.\,\ref{fig:4}B). In contrast, most other sensory contexts lead to numerous misclassifications, with some fish behaving more similarly to fish that belong to other sensory contexts.


\subsection*{Revealing phenotypic groups from inter-fish variability}

\paragraph{}
Our results point to hidden structure in the inter-individual variability that we now reveal with an unsupervised approach. We directly estimate the difference between fish through a modified Manhattan distance among the behavioral phenotypes (encoded in the transition matrices $T_q^f$ at $q=7$, see Methods) represented as a distance matrix $D_q$ (Fig.\,\ref{fig:5}A). To reveal significant structures in the phenotypic variability, we introduce a novel top-down clustering approach called Hierarchical multiplicative diffusive (HMD) clustering. The finite length of the recordings imposes an effective uncertainty on the measured transition matrices, as two distinct behavioral sequences  may result from a finite sampling of the same underlying transition matrix $T_q^f$. We estimate an effective significance scale for the behavioral phenotype of each fish $\hat{\epsilon}_{f}$ by leveraging simulations of symbolic sequences from $T_q^f$, which captures the inherent uncertainty over the precise location of each fish in the phenotypic space due to the finite size of the recordings (see Methods for details). Fig.\,\ref{fig:5}B represents a low-dimensional projection of the phenotypic space obtained using Constant Shift Embedding (CSE, \cite{roth2003optimal}): each gray point corresponds to a fish, and two example $q=7$ transition matrices are depicted in distinct positions of the space (red spots) with the significance scale $\hat{\epsilon}_{f}$ in blue, showcasing how neighboring fish may indeed be behave indistinguishably from each other within the finite size of the recording. Given this effective scale separation among fish, we search for an energy barrier on a family of multiplicative diffusion processes. We do this operation in an iterative fashion, performing a soft clustering along the highest effective energy barrier.  The outcome is the probability of each fish belonging to a given phenotypic group $\mathbb{G}$, $p(\mathbb{G}_i^L)$ at each iteration $L$. We stop the top-down subdivisioning when the gain in scale separation obtained after subdividing the space becomes negligible (Fig.\,\ref{Suppl:5}B). In Fig.\,\ref{fig:5}C, we illustrate the clustering outcome as a tree diagram, and color-code each fish in the transition matrix space according to their phenotypic group $\mathbb{G}_i, i \in \{1,\ldots,7\}$.

\paragraph{}
The phenotypic groups reflect the preferences for different motor strategies, Fig.\,\ref{fig:5}D. The main behavioral difference emerging from the top-down clustering is the use of fast cruising for $\mathbb{G}_{1,2,3}$. Conversely, fish in groups $\mathbb{G}_{4,5,6,7}$ opt for slow cruising and wandering behaviors. Further iterations reveal finer scale preferences for motor strategies. In Fig.\,\ref{Suppl:5}C we also report mean dwell times of fish conditioned on the group $\mathbb{G}$ they belong to. To further assess the role of preference in motor strategies, we evaluate how they lead to a differential exploration of space. We proceed as in Fig.\,\ref{fig:3}(B), and simulate artificial fish trajectories in each behavioral group using group level transition matrices $T^g$ (see Methods). We find that groups $\mathbb{G}_7$ and $\mathbb{G}_5$ are most efficient at pursuing nearby prey (Fig.\,\ref{fig:5}E1). Along with $\mathbb{G}_6$, these groups are also efficient at searching for resources uniformly scattered on distances up to $\approx 10$ body lengths (Fig.\,\ref{fig:5}E2). In contrast, group $\mathbb{G}_2$ is optimized for long distance dispersal, becoming most efficient on distances beyond $\approx 30$ body lengths. Notably, we also find that group $\mathbb{G}_{1,3,4}$ display average efficiency for searching at short length scales. At longer length scales $\mathbb{G}_{1,3}$ become more efficient, while $\mathbb{G}_4$ drops. We thus discover that the structure of the phenotypic space reflects an exploration-exploitation trade-off taking place at the level of the population.

\subsection*{Phenotypic groups capture behavioral differences imposed by sensory context}

\begin{figure*}[ht]
\centering
\includegraphics[width = 1.\textwidth]{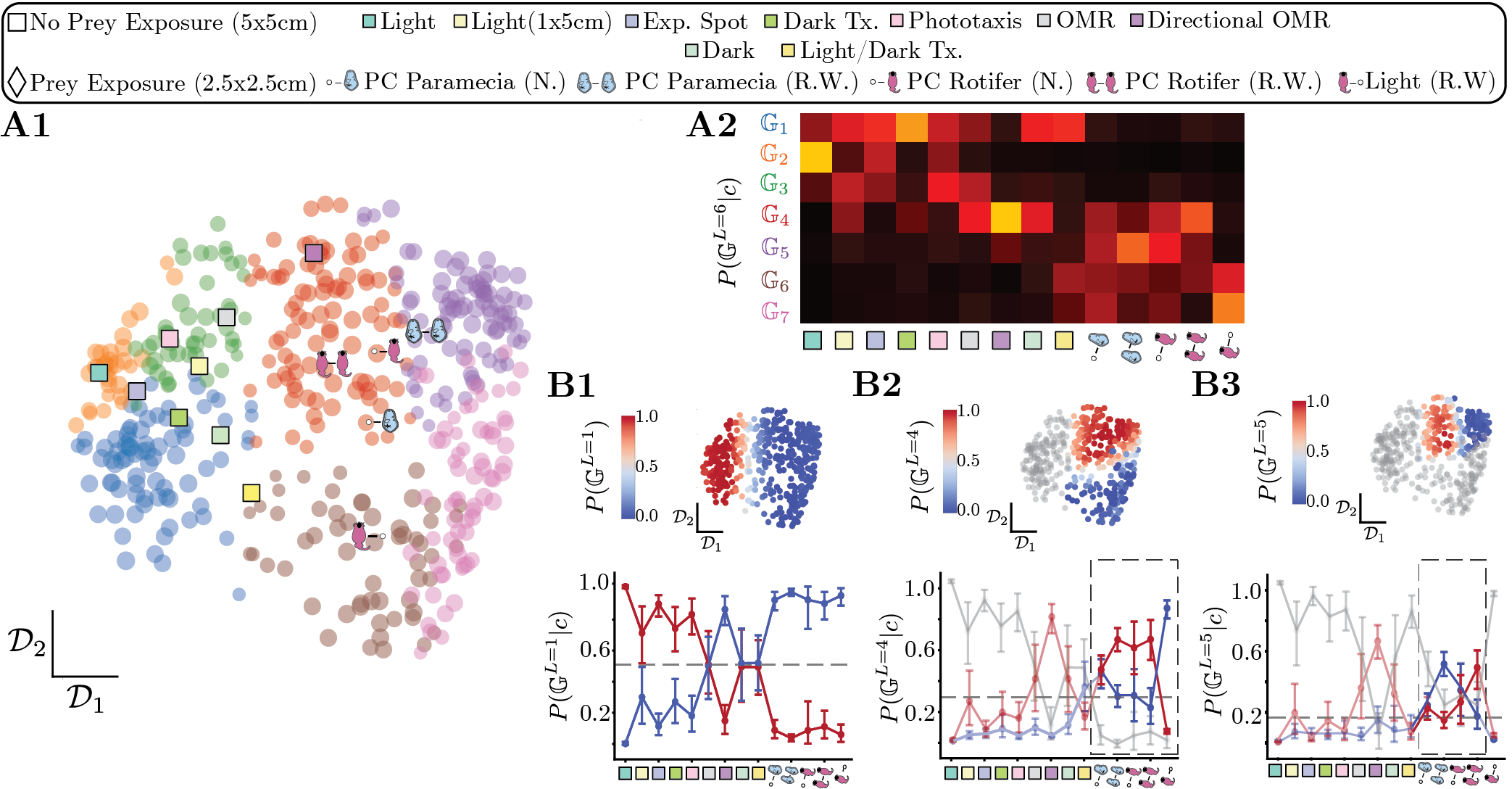}
\caption{\textbf{Phenotypic group structure explained by the sensory context, revealing prey exposure as a major determinant of variation} 
\textbf{(A1)} Transition matrix space color-coded by the 7 phenotypic groups. We also show the position of the average transition matrix in each behavioral assay, using the same markers as in Fig.\,\ref{fig:4}.
\textbf{(A2)} Probability of belonging to group $\mathbb{G}$ for fish in a given experimental condition $c$, $P(\mathbb{G}^{L=6}|c)$. Note that while no single sensory context maps onto a phenotypic group completely, the majority fish in certain contexts have preferences for certain groups. For example, most fish never exposed to prey belong to $\mathbb{G}_{1,2,3}$, while fish exposed to prey belong to groups $\mathbb{G}_{4,5,6,7}$
\textbf{(B)} To delineate this further, we show the probability of belonging to group $\mathbb{G}$ for fish in a given experimental condition $c$ at different iterations in the hierarchical subdivision $L$. The gray line represents the probability of belonging to neither of the subdivided groups.
\textbf{(B1)} The first split $L=1$ neatly distinguishes fish that were exposed to prey (either during or prior to the assay) from fish that were never exposed to prey. Notice also that the directional OMR condition is closer to the prey capture conditions, indicating that the sensory contexts in this assay somewhat emulate the spontaneous behavior in the prey capture assays. The OMR, Dark and Light/Dark transitions assays are distributed across the two main groups.
\textbf{(B2)} At the 4th iteration, $L=4$ fish that were previously exposed to prey but are in freely-swimming conditions neatly split from the fish that were in a hunting assay.
\textbf{(B3)} At the 5th iteration, $L=5$ fish in hunting assays that were raised with different types of prey mostly belong to different groups. Interestingly, naive fish that are hunting for the first time split equally, whereas fish that were raised with either paramecia or rotifers mostly belong to distinct groups.
}
\label{fig:6}
\end{figure*}

We now assess how much of the structure in the phenotypic space can be explained by sensory context. We place the average phenotype of each sensory context on the phenotypic space (Fig.\,\ref{fig:6}A1), and estimate how many fish in a given sensory context belong to each behavioral group in Fig.\,\ref{fig:6}A2. We find that sensory contexts can drive preferences for different regions of the phenotypic space (Fig.\,\ref{fig:6}A2), but no single sensory context completely maps onto a single phenotypic group. 
 
\paragraph{}
To elucidate this further, we evaluate how sensory contexts split at different levels of the hierarchical clustering process. Fig.\,\ref{fig:6}B1-B3 shows the probability of belonging to a given group $\mathbb{G}_i^L$ at different subdivision levels $L$. The first iteration $L=1$ captures the main axis of variation by differentiating naive fish from fish exposed to prey (Fig.\,\ref{fig:6}B1). The largest phenotypic variation favoring high exploitation emerges upon exposure to prey, changing the behavior to promote wandering and slow cruising. At $L=4$ iterations, the phenotypic groups $\mathbb{G}_{4,5}$ differ from $\mathbb{G}_{6,7}$ (Fig.\,\ref{fig:6}B2), thereby separating fish raised with rotifers but freely exploring without prey (Fig.\,\ref{fig:6}B2 below in blue) from fish hunting (Fig.\,\ref{fig:6}B2 below in red). A prior hunting experience impacts also the freely swimming behavior of the fish, resulting in a dynamic phenotype that is different not just from naive fish but also from fish in hunting assays (Fig.\,\ref{fig:6}B2). At the $L=5$ iteration highlight differences between fish hunting for paramecia versus rotifers (Fig.\,\ref{fig:6}B3). The $L=5$ iteration highlights differences between fish hunting for paramecia versus rotifers (Fig.\,\ref{fig:6}B3), which become more evident upon raising the fish with either prey type from 3dpf. Altogether, our approach uncovers how different prior exposure to prey leads to changing hunting phenotypes at a later developmental stage. 

\paragraph{}
To further assess the differences in behavior among naive fish and those raised with prey, we quantify the spatial distributions of different motor while taking into account the effects of the arena geometry (Fig.\,\ref{Suppl:6}A,B). In freely-exploring conditions, naive fish wander mostly only at the corners of the arena when forced to reorient (Fig.\,\ref{Suppl:6}A,B) -- both in the $5\,\text{cm}\times 5\,\text{cm}$ and in the $1\,\text{cm}\times 5\,\text{cm}$ arenas. In contrast, fish raised with prey and freely exploring in a $2.5\,\text{cm}\times 2.5\,\text{cm}$ arena performed wandering throughout the entire arena (Fig.\,\ref{Suppl:6}C,D) and were more frequently oriented towards the wall, suggesting that wandering is used as a local search strategy to leave the arena (\cite{fero2011behavioral, schnorr2012measuring}).

\subsection*{Phenotypic groups reveal persistent hidden states impacting behavior across timescales}

\begin{figure*}[ht]
\centering
\includegraphics[width = 1.\textwidth]{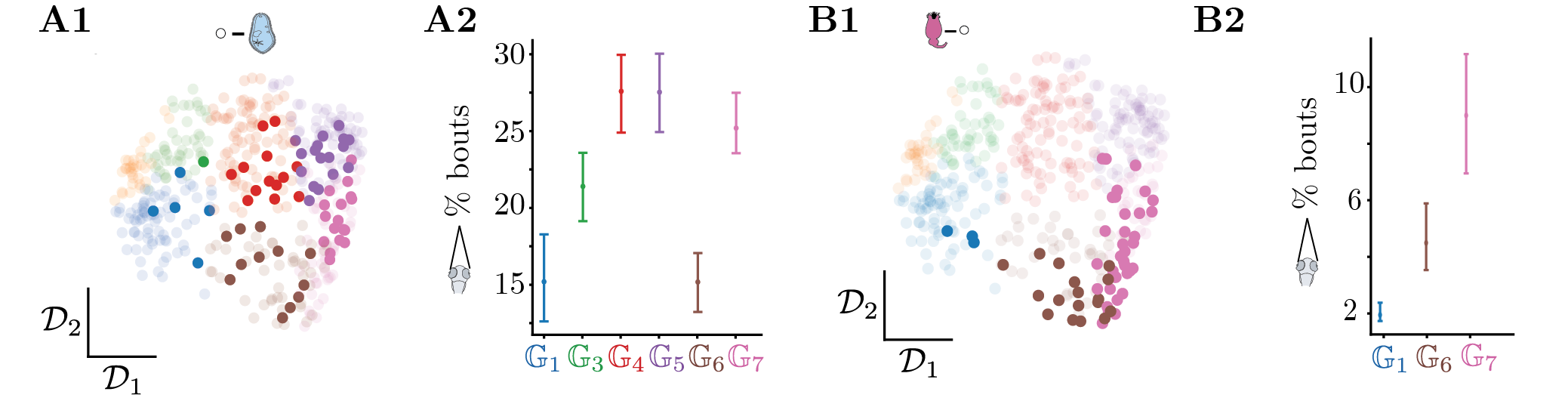}
\caption{\textbf{Phenotypic group structure reveals differences in sensorimotor transformations among fish within a sensory context}
\textbf{(A1)} Transition matrix space color-coded by the 7 phenotypic groups, of naive fish hunting for paramecia (represented in bold). Note the high degree of variability in the dynamics of the fish within this condition.
\textbf{(A2)} Percentage bouts spent in eye convergence throughout the experiment for naive fish hunting paramecia, structured by the phenotypic group $\mathbb{G}$. Note that fish in $\mathbb{G}_1$ and $\mathbb{G}_6$ have lower time spent with eyes converged, indicating that these fish are potentially hunting lesser than fish in other groups.
\textbf{(B1)} Transition matrix space color-coded by the 7 phenotypic groups, of fish freely exploring in the light, but raised with rotifers (represented in bold). 
\textbf{(B2)} Percentage time spent in eye convergence across fish in the light raised with condition, structured by the phenotypic group $\mathbb{G}$. Note that while the overall time spent in eye convergence is quite low as these fish are freely swimming, fish in $\mathbb{G}_7$ still have significantly higher time in eye convergence, pointing to differing hidden states impacting behavior within this condition.
}
\label{fig:7}
\end{figure*}

Sensory contexts captures some of the variation in the phenotypic space, but we also find a large amount of variability within sensory contexts (Fig.\,\ref{Suppl:7}), pointing to hidden variables that impact the sensorimotor transformations performed by the fish. We therefore investigate whether such variability can reflect inner motivational states by investigating the variability observed upon exposure to prey. We focus on fish hunting for paramecia (bold highlight, Fig.\,\ref{fig:7}). Half the fish in this assay do not belong to the population average groups $\mathbb{G}_4$ and $\mathbb{G}_5$. As a hunting indicator, we leverage the measurement of time spent with eye converged throughout the experiment (\cite{bianco2011prey}). Fish belonging to exploration phenotypes ($\mathbb{G}_{1,3}$) rarely hunt, whereas fish whose phenotype is tuned to pursuing prey ($\mathbb{G}_{5,7}$) hunt often (Fig.\,\ref{fig:7}A2). Interestingly, fish raised with rotifers but freely swimming without prey in the light (Fig.\,\ref{fig:7}B1) also show differences in their eye convergence rates that depend on the phenotypic groups: fish tuned to exploration ($\mathbb{G}_{1}$) barely attempt hunts, whereas fish that are more tuned to exploit ($\mathbb{G}_{7}$) exhibited a higher rate of eye convergence (Fig.\,\ref{fig:7}B2). Overall, while our approach does not directly take into account the prey capture rate, the structure to the variability reveals evidence for motivational states that drive contrasting exploration-exploitation phenotypes.
 
\label{s:results}

\section*{Discussion}

In this study, we aim to reveal how behavioral phenotypes arise as a function of sensory inputs and latent variables that impact behavior on multiple timescales. Via the analysis of maximally-predictive bout sequences in larval zebrafish, we reveal a hierarchy of three long-lived modes of navigation organized by timescale: a long lasting mode that reflects the rate of reorientation, a faster mode that also encodes for speed and a third mode that captures egocentric direction preferences. We then discover phenotypic preferences along these modes among individuals which seem to be partially explained by the sensory context (arena size, illumination and experimental stimuli). To quantify the structure and origin of this variation, we compare individual fish to reveal phenotypic groups using only the behavior at the experimental timescale. Through \emph{in silico} experiments we find that the phenotypic group structure corresponds to exploration v/s exploitation trade-offs apparent across individuals. These groups reveal that a major driver of phenotypic variation is the exposure to prey, engaging a strong preference for local exploitation. This variation not only impacts the dynamics deployed by animals in hunting assays but also affects the freely swimming behavior of fish in the light. However, the structure of the phenotypic space also reveals similarities between fish in distinct sensory contexts, pointing to a combined impact of sensory context and persistent motivational states that differ among individuals.  We show that phenotypic group structure conditions the sensorimotor transformation at short timescales by promoting either exploitation or exploration, showcasing how these latent variables prevail over the immediate sensory contexts.


\subsection*{A hierarchical organization of timescales in larval zebrafish behavior}

Sensory-evoked navigation consists in chaining locomotor bouts in response to external sensory cues from the environment and internal states. We reveal long-lived modes of behavior by constructing Markov models from bout sequences deployed by the larval zebrafish, accounting for the history dependence in the behavior and the maximal possible variability in posture dynamics of bouts (\cite{costa2023maximally}). Our method consistently provided a hierarchy of motor strategies, reflecting rate of reorientation, speed and directionality, in distinct datasets acquired by different users, from different arenas, and laboratories. 
Changes in long-lived modes corroborate observations from previous studies noticing changes in reorientation rates in the dark (\cite{horstick2017search}), speed modulation during optomotor response (\cite{severi2014neural}) or recurrence in left or right bias during navigation in the light (\cite{dunn2016brain}) or in response to the dark (\cite{horstick2017search}; cite Hageter et al., Frontiers in Behavioral Neuroscience 2021). However, these modulations were previously observed by focusing on a specific kinematic parameters, choosing a specific time-window to analyze, or classifying all bouts into a limited repertoire (such as left versus right for (\cite{dunn2016brain}). Our unbiased analysis expands this view by quantifying the dynamics of behavior as the chaining rule of maximally predictive bout sequences, thereby providing a complete and simultaneous picture of the multiple axes of behavior and revealing how they are organized hierarchically by timescale, which could have been missed upon focusing on specific kinematic parameters.


\paragraph{}
Methodologically, we use the inferred long-lived modes of a transfer operator to effectively identify an increasing number of motor strategies that capture shorter lived behaviors (Fig.\,\ref{fig:2}). Coarse-graining behavior using the non-trivial eigenvectors of transfer operators is equivalent to using the Information Bottleneck approach (\cite{schmitt2023information}), commonly applied in behavioral analysis (\cite{berman2016predictability}). This equivalence holds for behavioral representations that evolve in a Markovian fashion, stressing the importance of working with maximally predictive bout sequences (\cite{costa2023maximally,costa2023markovian}). Additionally, it is common to define discrete states by clustering similar movements, identifying movements that are executed often, or by building Hidden Markov Models (HMM) (\cite{marques2018structure,johnson2020probabilistic,mearns2020deconstructing,wiltschko2015mapping,wiltschko2020revealing,calhoun2019unsupervised}). However, the level of discreteness required is challenging to define and interpret with such approaches. To avoid this issue, we use the notion of timescale separation as a guiding principle for defining when a given discrete representation holds, which we reveal by coarse-graining according to the \emph{dynamics}. On the longest timescales, cruising-wandering strategies provide a good-enough representation. To capture faster timescale behaviors, we utilize a increasing number of finer scale motor strategies. From these motor strategies, we build transition matrices for individual fish that are predictive of each fish's behavior across coarse-graining scales. This parsimonious encoding of the behavioral program of each fish quantifies its behavioral phenotype, providing us with specific timescales in the motor strategies deployed by individuals. If these motor strategies were arbitrarily defined, simple Markov models would not make good predictions. 


\paragraph{}
The predictive power of our approach is adequate but has some limitations: while we can predict the probability that the next bout sequence will belong to a coarse-grained motor strategy, we cannot predict the precise kinematics of each bout in the sequence. This may simply reflect an inherent bound to predictability coming from the stochastic nature of the behavioral dynamics, or be due to insufficient data. Further work will help resolve this question. While our approach is effective at capturing behavioral dynamics across several timescales, there are opportunities for enhancement: i) longer behavioral recordings in larger arenas with novel sensory contexts will likely reveal new long-lived behaviors;  ii) we could include the spatiotemporal properties of the stimuli and the arena geometry, which impacts the dispersal properties of the fish's behavior (Fig.\,\ref{Suppl:8}); iii) to capture any slow non-stationary changes to behavior, we could introduce time-varying transition rates as was done in  \cite{costa2024fluctuating}; iv) we could incorporate the inter-bout intervals in our approach as we find them correlated with the bout sequences used (Fig.\,\ref{Suppl:2} h1-h3), similar to a previous observation (\cite{johnson2020probabilistic}).

\subsection*{Sensory contexts drive overall preferences for motor strategies}

Behavior in larval zebrafish is classically studied to investigate instantaneous sensorimotor transformations. By quantifying the behavioral responses of a large population of fish, our approach uncovered biases of motor strategies across multiple timescales in response to diverse sensory contexts. We show that larval zebrafish display a preference for fast cruising when freely swimming in the light, while in aversive settings they display preferences for wandering. Previous work discovered that upon exposure to darkness, larval zebrafish  spiral for 2-3 minutes as a local search and turn more in the dark (\cite{horstick2017search}). Our results confirm these observations, revealing long-lived wandering states in the dark ($\approx 8\,\text{s}$) interspersed by short segments of cruising ($\approx 4\,\text{s}$), so that fish mostly wander in dark environments ($\approx 30\,\text{min}$), (Fig.\,\ref{fig:3}A3). Our simulations indicate that higher wandering corresponds to the fish performing more local area searches. The intermittent switches to cruising could help the fish to disperse to newer areas faster, as previously hypothesized (\cite{horstick2017search}). 

\paragraph{}
Recent studies have identified stereotyped bout types used during hunting sequences through unsupervised clustering (\cite{marques2018structure,johnson2020probabilistic,mearns2020deconstructing}), and analysis of fish dispersal within pre-selected time-windows enabled the discovery of long-lived ``exploration'' and ``exploitation'' states in hunting assays (\cite{marques2020internal}). We find that the hunting sequence is represented in slow cruising, which constrains the variation in cruising timescales (Fig.\,\ref{fig:3}A4). In the inter-hunt period, fish mostly engage in wandering strategies which could correspond to fish actively searching for prey in their surroundings. Thus beyond the tight stimulus-response loop of the hunting sequence, the behavior of the fish throughout the experiment is overall attuned towards searching and hunting. Furthermore, our comparative analysis over multiple sensory contexts differentiates the inter-hunt exploratory period from navigation in the light, during which fish mostly display fast cruising. 
Fast cruising may thus represent an alternative ``exploration'' strategy that is extremely rare upon exposure to prey. In \emph{C. elegans}, exploration-exploitation behaviors have been linked to minutes long ``roaming'' and ``dwelling'' states respectively (\cite{fujiwara2002regulation,flavell2013serotonin}): the ``roaming'' state is dominated by faster forward locomotion with rare reorientations, while the ``dwelling'' state is characterized by smaller scale movements that do not coherently engage the whole body resulting in low speeds \cite{fujiwara2002regulation,flavell2013serotonin,hebert2021wormpose}. We hypothesize that ``slow cruising and wandering'', with its slow speeds and targeted reorientations may be analogous to the ``dwelling'' state of \emph{C. elegans}, while ``roaming'' recalls ``fast cruising'' behavior, which is also dominated by forward bouts and sporadic routine turns and is most efficient at long distance dispersal. Our approach offers an opportunity to further dissect the neuromodulatory control mechanisms driving these exploration-exploitation trade-offs by taking into account the finer scale bout sequence dynamics, thus acting as a powerful complement to previous approaches that quantified the overall displacement of the fish within pre-selected time-windows (\cite{marques2020internal}). 

\paragraph{}
By providing a simultaneous handle on the multiple timescales that govern behavior such as cruising and wandering, differential modulation of speed and egocentric direction preference, our approach sets the stage for dissecting the underlying brain circuits for navigation in a transparent vertebrate brain. Along with the serotoninergic modulation of motor activity mentioned previously, we can dissect how multiple brain regions function together to give rise to this hierarchy of timescales in behavior. The hindbrain oscillator, also called anterior rhombencephalic turning region  has been identified as one brain region that confers a persistence of the left/right steering in the larval zebrafish (\cite{dunn2016brain, wolf2017sensorimotor}). We hypothesize that the interplay of this region with the mesencephalic locomotor region (MLR, \cite{carbo2023mesencephalic}), the nucleus of the medial longitudinal fasciculus  (\cite{wang2014selective,berg2023brainstem}), projecting onto the reticular formation (\cite{orger2008control,carbo2023mesencephalic}) could explain the dynamics we uncover. In particular, the modulation of the MLR by dopamine released from posterior tuberculum (\cite{carbo2023mesencephalic}) could explain the long-lived persistence of cruising behaviors. There is already some evidence for this in mice, where dopamine release has been associated with long time scale (seconds to minutes) persistence of motor sequences (\cite{markowitz2023spontaneous}). 

\subsection*{Variability in multiscale behaviors exhibits hierarchical structure}

While it is apparent that sensory contexts give rise to preferences for motor strategies, it is not clear how these contexts shape the structure of phenotypic variation. This is further confounded by the large amount of individual variability, evidenced in our inability to predict the sensory context of many individuals based on their behavior alone (Fig.\ref{fig:4}). We hypothesize that the variability in emergent phenotypes is driven by a combination of sensory contexts and persistent hidden states. To study the structure of this variability, we introduce an unbiased approach that compares individual animals directly using their multiscale behavior to reveal phenotypic groups. Recent studies developed strong insight into the structure of behavioral variability by comparing important kinematic parameters (\cite{werkhoven2021structure}), the probability of the occurrence of behaviors (\cite{hernandez2021framework}), or the parameters of minimal behavioral models (\cite{helms2019modelling,goc2021thermal}). Our approach for comparing individuals incorporate all of these aspects: 1) the hierarchy of timescales in the ensemble dynamics naturally reveals the important behavioral parameters in the form of interpretable motor strategies; 2) the predictive power of our coarse-grained Markov models allows us to encode the dynamics of the behavioral transitions for each individual, capturing also the probability of visiting each behavior; 3) our ability to scan across coarse-graining scales reveals the level of fine-scale information most adequate for comparing animals.

\paragraph{}
Quantifying structure in individuality is a non-trivial task due to the uncertainty associated with the variable duration and fish swimming frequency across recordings. This uncertainty renders bottom-up agglomerative clustering approaches unreliable. We solved this issue by turning this limitation into a feature, introducing a novel top-down clustering algorithm that directly leverages the uncertainty in the estimate of each behavioral phenotype to provide an effective scale separation between individuals. While this clustering method was developed for analyzing behavior in freely moving animals, it could be generalized to the study of other dynamic biological processes where variability around common principles is the hallmark, such as cell migration (\cite{bruckner2020disentangling}) or neural dynamics (\cite{brynildsen2023network}).

\subsection*{Behavioral phenotypic groups emerge as consequence of both sensory drives and persistent hidden states}

Prey capture is an innate behavior as larval zebrafish at 6-7 dpf must feed to ensure their survival (\cite{wilson2012aspects}). Previous exposure to prey has been shown to increase their rate of capture initiation (\cite{oldfield2020experience,lagogiannis2020learning}). The first split in the top-down clustering process is largely driven by exposure to prey (Fig.\,\ref{fig:6}A1): most fish exposed to prey perform local exploitation ($\mathbb{G}_{4,5,6,7}$) while naive fish perform long distance dispersal ($\mathbb{G}_{1,2,3}$). This exposure to prey has a even stronger impact on the behavioral phenotype of fish than aversion to darkness (Fig.\,\ref{Suppl:7}). Remarkably, even fish previously raised with prey but recorded freely exploring in the light belong to exploitation groups ($\mathbb{G}_{6,7}$).  Such long lasting preference may be due to recruitment of the hypothalamus during hunting, initiating a feeding state in the fish (\cite{muto2017activation}). Serotoninergic (\cite{filosa2016feeding}) and dopaminergic signalling (\cite{zaupa2024calmodulin}) may also be implicated in extending duration of this feeding state, manifesting as a large phenotypic variation in behavior. We also find that the nature of the preys (paramecia or rotifers, \cite{marques2018structure})) may lead to distinct behavioral phenotypes suggesting that fish can adapt their motor strategies to the kinematics of the prey. Our work opens the possibility for future studies investigating the dynamics of predator-prey interactions in greater detail.

\paragraph{}
While the exposure to prey induces exploitation-related phenotypes, we also find a significant number of fish belonging to exploration phenotypes (and vice versa for fish never exposed to prey). One can speculate on what these hidden states may correspond to. Fish in hunting assays that deviate from the population average also display lower rates of eye convergence, possibly due to different motivational states such as feeding or arousal that can profoundly impact sensorimotor integration (\cite{pantoja2016neuromodulatory,pantoja2020rapid}).

\paragraph{}
Our work provides novel approaches for the quantitative study of phenotypic variability across timescales. We reveal how this variability is structured by the sensory context of the animal and hidden motivational states that drive the animals along an exploration-exploitation trade-off. Our approach lays the groundwork for the study of the sources of individual variability, and can be easily extended to other species. Combined with recent advances in large scale tracking of wild animals, we offer a new avenue to study how phenotypic variation is shaped by the environment, serving the cause of ecology and biodiversity. 

\label{s:discussion}

\section*{Methods}

\subsection*{Software availability} Code and data for reproducing our results is publicly available at \url{https://github.com/GautamSridhar/Markov_Fish}

\subsection*{Preprocessing of the data from \cite{marques2018structure}}

In the dataset from \cite{marques2018structure}, wild-type Tubingen zebrafish larvae (\textit{Danio rerio}) were used at 6-7 days post fertilization. Larvae were recorded at a high temporal resolution (700Hz), at two different pixel sizes ($58\mu\text{m}$ in the $5\times 5\,\text{cm}^2$ arenas and 27$\mu$m in the $2.5\times 2.5\,\text{cm}^2$ arenas) and tail-tracking of 8 points on the tail was performed online using custom software (\cite{marques2018structure}). From this dataset, we collect 463 fish from 14 varying sensory contexts (Table \,\ref{Suppl:Table}). In \cite{marques2018structure}, we retained all detected bouts that generated a velocity of at least $4\,\text{mm/s}$ (equivalent to one body length of the fish). We also noticed that the detected bout start and bout end cut-off were slightly inaccurate and many detected bouts did not showcase the full motion of the tail from start to end. To account for this, we took 10 extra frames from before the start of the bout as the new bout start frame. For the new bout end, we chose the frame up to when the fish was generating a velocity of at least 4mm/s (equivalent to 1 body length), or 175 frames (equivalent to 250ms), whichever limit came first. We provide further details for each behavioral assay below:
\vspace{0.1cm}

{\bf Light }$\mathbf{5\times 5\,\text{cm}^2}$: Fish were presented with a uniform light from below with an illuminance of $1000\,\text{lm/m}^2$ in a $5\times 5\,\text{cm}^2$ squared arena with $3\,\text{mm}$ of depth. 
\vspace{0.1cm}

{\bf Light }$\mathbf{1\times 5\,\text{cm}^2}$: Fish were presented with a uniform light from below with an illuminance of $1000\,\text{lm/m}^2$ in a $1\times 5\,\text{cm}^2$ squared arena with $8\,\text{mm}$ of depth. 
\vspace{0.1cm}

{\bf Dark }$\mathbf{5\times 5\,\text{cm}^2}$: Fish were presented with darkness ($0\,\text{lm/m}^2$) in a $5\times 5\,\text{cm}^2$ squared arena with $3\,\text{mm}$ of depth. 
\vspace{0.1cm}

{\bf Expanding Spot }$\mathbf{5\times 5\,\text{cm}^2}$: An expanding dark spot at different speeds (0.25, 0.5, 1, 1.5, 2.0, 2.5 cm/s) and different orientations ($0^\circ$, $90^\circ$, $180^\circ$, $270^\circ$) were presented in closed loop 4 cm away from the larva. Stimuli were randomized and presented every 2 min. This assay was done in a 5cm x 5cm squared arena with 0.3 cm of depth.
\vspace{0.1cm}

{\bf Dark Transitions }$\mathbf{5\times 5\,\text{cm}^2}$: The spontaneous swimming with light transitions assay was based on \cite{burgess2007modulation}. Fish were left in the dark for 30 minutes and then presented with uniform light at different intensities (0, 12, 44, 104, 232, 447, 790, 1890, 4700 $\,\text{lm/m}^2$) for 3 min in a $5\times 5\,\text{cm}^2$ arena with $3\,\text{mm}$ of depth.  
\vspace{0.1cm}

{\bf Phototaxis }$\mathbf{5\times 5\,\text{cm}^2}$: The phototaxis assay was based on \cite{huang2013spinal} and performed in $5\times 5\,\text{cm}^2$ arena with $3\,\text{mm}$ of depth. The stimulus consisted on a uniform brightness of varying intensity (100, 410, 780, 1250 $\,\text{lm/m}^2$) on one side of the fish and darkness on the other. The stimulus was in closed loop with the larva for a duration of $12\,\text{s}$. 
\vspace{0.1cm}

{\bf Forward Optomotor response }$\mathbf{1\times 5\,\text{cm}^2}$: The forward optomotor response assay was performed as described in \cite{severi2014neural}, but using a $1\times 5\,\text{cm}^2$ with $8\,\text{mm}$ depth arena. Drifting gratings with a spatial period of $1\,\text{cm}$ and ten different speeds (0, 2.5, 5, 7.5, 10, 15, 20, 30, 40, 50 mm/s) were presented from below when the larva was at the extremities of the arena. Trials would end when fish reached the opposite end of the arena. After, there were 5s of intertrial interval of homogenous light ($1000\,\text{lm/m}^2$) and a new trial would start with gratings in the opposite direction. In trials that larvae were not able to reach the opposite end of the arena ($> 30\,\text{s}$) a $10\,\text{mm/s}$ grating was displayed until it swam the remaining distance. 
\vspace{0.1cm}

{\bf Directional Optomotor response }$\mathbf{5\times 5\,\text{cm}^2}$: The directional optomotor response assay was performed as in \cite{orger2008control}. A $\mathbf{5\times 5\,\text{cm}^2}$ arena with $3\,\text{mm}$ of depth was used. Drifting gratings with a spatial period of $1\,\text{cm}$, moving at $10\,\text{mm/s}$, and from 24 different orientations ($15^\circ$ apart) were presented from bellow to the larva and in closed loop with its orientation. Trials lasted 10s and started when the larva was in the center of the arena. During the inter-trial interval, that would last at least 5s, circular converging gratings were projected to drive the larva to center of the arena. 
\vspace{0.1cm}

{\bf High Lux light-dark transitions }$\mathbf{5\times 5\,\text{cm}^2}$: Fish were exposed to alternating 3 min periods of high illuminance light ($5000\,\text{lm/m}^2$) and darkness in an $5\times 5\,\text{cm}^2$ arena with $3\,\text{mm}$ of depth.  
\vspace{0.1cm}

{\bf Prey Capture }$\mathbf{2.5\times 2.5\,\text{cm}^2}$: All prey capture datasets were performed in $2.5\times 2.5\,\text{cm}^2$ arenas with $3\,\text{mm}$ depth and larvae were illuminated from above ($1000\,\text{lm/m}^2$). Fish were fed with 50-100 paramecia (\emph{Paramecium caudatum}) or rotifers (\emph{Brachionus plicatilis}) and were allowed to hunt for 1-2h. A subset of fish were never fed until the assay, while others were raised, starting at 3 days post fertilization, with the type of prey that was used in the assay. To ensure that these fish were not satiated they were starved for at least 2 hours before the assay. 
\vspace{0.1cm}

{\bf Freely exploring in the light, reared with rotifers }$\mathbf{2.5\times 2.5\,\text{cm}^2}$: Fish were fed with 50-100 rotifers starting at 3 days post fertilization and then placed in $2.5\times 2.5\,\text{cm}^2$ arenas with $3\,\text{mm}$ depth and illuminated from above ($1000\,\text{lm/m}^2$) to freely swim. To ensure that these fish were not satiated they were starved for at least 2 hours before the assay. 

\subsection*{Data collection and preprocessing of the data from \cite{reddy2022lexical}}

In the dataset from \cite{reddy2022lexical}, wild-type AB zebrafish larvae are recorded at 7 days post fertilization. Larvae were placed in arenas of size $14\,\text{cm}\times 1\,\text{cm}\times 0.4\,\text{cm}$ and recorded for 10 minutes at a temporal resolution of $160\,\text{Hz}$ and pixel size of $70\mu\text{m}$. Half the fish in each experiment were exposed to acidic ($\text{pH}\approx 2$) gradients at both horizontal ends of the arena. After diffusion, the acidic solution formed a steep gradient $20\,\text{mm}$ away from each end of the arena. 8 points on the fish's tail were tracked using Zebrazoom (\href{https://zebrazoom.org/}{https://zebrazoom.org/}) and recordings were processed as indicated in \cite{reddy2022lexical}. We retain 218 fish in this dataset which performed more than 250 bouts during the recording. Similar to the previous dataset, we keep all bouts up to either the detected bout ends, at most till 40 frames (250ms).

\subsection*{Principle component analysis (PCA) on the space of bouts}
By measuring 8 tail angles over $N_\text{frames}$ (175 frames for the data from \cite{marques2018structure} and 40 frames for the data from \cite{reddy2022lexical}), we end up with a high dimensional $N_\text{frames} \times 8$ representation of each bout. To increase tractability while reducing noise, we perform a principle component analysis (PCA), Fig.\,\ref{Suppl:1}. For the dataset of \cite{marques2018structure}, we randomly sample 25 recordings 180 times (amounting to $\approx 50,000$ bouts per sample) and then calculate the eigenvectors and eigenvalues of the covariance matrix built along the bouts of this recording. Across multiple resamples of the data, we received very stable eigenvalues as represented by the miniscule 95\% errorbars in Fig.\,\ref{Suppl:1}A. We then average the covariance matrices across these multiple resamples and use the eigenvectors of the averaged covariance matrix as the eigenvectors to represent the feature space of bouts (Fig.\,\ref{Suppl:1}B). We retain 20 dimensions for the dataset from \cite{marques2018structure} which covers upwards of 95\% of the cumulative variance explained (\textbf{Fig.\ref{Suppl:1}a}). Similarly, for the dataset from \cite{reddy2022lexical}, we retain 12 principle components.

\subsection*{Maximally predictive state spaces for larval zebrafish behavior}
To study the dynamics in the bout space, we adapt the method described in \cite{costa2023maximally, costa2023markovian}. Given $l$ bouts (where one bout is a $d$ dimensional object after PCA) in a recording, we stack $K$ bouts together in overlapping windows to receive a matrix of size $(l-K+1) \times dK$. This process is repeated for multiple values of $K$. In order to approximate the dynamics in a high dimensional delay-embedding space,  we rely on a discrete approximation of the transfer operator as described in \cite{costa2023maximally} - we first partition the bout sequence space through \emph{k-means} clustering for which we utilize the scikit-learn function with \emph{k-means ++} initialization (\cite{scikit-learn}). After partitioning the space into $N$ microstates, the dynamics is recapitulated through a transition matrix $T_\text{ensemble}$ that is built by sampling sequences of bouts equally from each sensory context. We then check for which value $K^*$ the short time entropy of rate of $T_\text{ensemble}$ minimizes. The entropy rate is measured as $h = -\sum_{ij}\pi_i T_{ij}\log T_{ij}$, where $\pi_i$ represents the $i$-th entry of the invariant density, obtained as the first eigenvector of $T_{ij}$ (see \cite{costa2023maximally} for further details). We then search for $K^*$ such that  f$\partial_K{h}(K^*) \sim 0$, Fig.\,\ref{fig:1}(d),\ref{Suppl:4}. We similarly select the number of partitions $N=N^*$ for the correct $K=K^*$ at the maximum point just before finite-size effects cause a reduction the short-time entropy rate (Fig.\,\ref{fig:1}(D),\ref{Suppl:4}). This allows us to incorporate as much information about the fine-scaled dynamics as possible. In this fashion, we have moved from a high-dimensional continuous space of bout evolution to a Markov chain representing the dynamics as a sequence of $N^*$ symbols.

\paragraph{}
For the datasets from \cite{marques2018structure}, we utilize $463$ fish from $14$ different sensory contexts (\textbf{Table \ref{Suppl:Table}}). We estimate the entropy rates by building transition matrices $T_\text{ensemble}$, and bootstrapping across a random sampling of 7500 bouts from each condition over 100 random seeds, to ensure a uniform sampling across sensory contexts. For the dataset from \cite{reddy2022lexical}, we have 2 sensory contexts. The entropy rate reflects a bootstrapping across a random sampling of 40,000 bouts from each condition over 50 random seeds. Once $K^*$ and $N^*$ were chosen, we selected the cluster labels that maximize the entropy rate with respect to the random resampling, and use such cluster labels for the rest of the analysis.

\subsection*{Building an ensemble transition matrix $T_\text{ensemble}$ }
To study the dynamics of maximally predictive bout sequences across fish, we estimate an ensemble transition matrix $T_\text{ensemble}$ from the available data. At $K^*=5$ and $N^*=1200$, we randomly sample 7500 bout sequences from each condition over 100 seeds and estimate $T_{\text{seed}}(\tau)$ for varying transition times $\tau$ for each seed. We estimate the implied timescales of the reversibilized transfer operator from the eigenvalues of each reversibilized $T_{\text{seed}}(\tau)$, $\lambda_i(\tau)$, $t^{\text{imp}}_i(\tau) = \frac{-\tau}{\log\lambda_i(\tau)}$, and choose $\tau^*=3\,\text{bouts}$ such that the fine-scaled dynamics have relaxed to the steady state distribution, isolating the long-lived modes (see \cite{costa2023maximally} for further details), Fig.\,\ref{Suppl:2}A. We calculate bootstrapped estimates of the implied timescales across the 100 estimates of $T_{\text{seed}}(\tau)$. The noise floor is calculated by shuffling the symbolic sequence of the dynamics, re-estimating reversibilized transition matrices and obtaining their largest real eigenvalue. Finally, we estimate the ensemble $T_\text{ensemble}$ by averaging the transition matrices $T_{\text{seed}}(\tau^*)$ and row-normalizing.

\subsection*{Operator based partitioning into metastable strategies}
The eigenvectors of the reversibilized transition matrix $T_\text{ensemble}$ can be used to provide an effective coarse-graining of the dynamics into metastable strategies via the formulation of almost-invariant sets (\cite{froyland2003detecting,froyland2005statistically,costa2023maximally}. Each eigenvector of the reversibilized $T_\text{ensemble}$ provides an ordering of the behavioural state space along a particular timescale. To find the first level of coarse-graining into the two metastable strategies of cruising-wandering, we rely on the fact that the first eigenvector provides an optimal 2-way cut of a graph encoding the dynamics (\cite{froyland2003detecting,froyland2005statistically}). In practice, we define two macroscopic sets (which correspond to collections of microstates $s_i$) by splitting along $\phi_1$,

\begin{equation*}
    S^+ (\phi_1^c) \coloneqq \bigcup_{i:\phi_1\geq \phi_1^c} s_i\,\\,\,S^- (\phi_1^c)\coloneqq \bigcup_{i:\phi_1<\phi_1^c} s_i,
\end{equation*}

\noindent where $\phi_1^c$ is a threshold that is chosen to maximize the metastability of a set. We measure the metastability of each set $S$ by estimating how much of the probability density remains in $S$ after a time scale $\tau$, 

\begin{equation*}\label{Eq:coherence}
    \chi_{\pi,\tau}(S) = \frac{\sum_{i,j\in S}\pi_i T_{ij}(\tau)}{\sum_{i \in S}  \pi_i},
\end{equation*}
where $\pi$ is the the invariant density, estimated directly from $T$. To estimate the overall measure of metastability across both sets $S^+$ and $S^-$, we define

\begin{equation}\label{Eq:coherence_min}
    \chi(\phi_1^c) = \min\left\{\chi_{\pi,\tau^*}(S^+),\chi_{\pi,\tau^*}(S^-)\right\}.
\end{equation}
which we maximize with respect to $\phi_1^c$. See \cite{costa2023maximally} for further details and applications to known dynamical systems. In Fig.\,\ref{Suppl:2}B we show the overall coherence measure as a function of $\phi_1$, which we use to find the $\phi_1^c$ that maximizes coherence, defining ``cruising'' and ``wandering'' states.

\paragraph{}
Subsequent coarse-graining into $q$ shorter timescale strategies is achieved by adapting the \emph{q-way} cut formulation from \cite{froyland2003detecting}, which relies on performing a \emph{k-means} clustering using the first $\lceil{\log_2 q}\rceil$ non-trivial eigenvectors of the reversibilized transition matrix. We introduce a new heuristic for obtaining \emph{q-way} cuts that respect the metastability of the dynamics. We first identify the value of each $\phi_k$ that maximizes the coherence Eq.\,\ref{Eq:coherence_min}, defining $S^+$ and $S^-$ by thresholding along each eigenvector $\phi_k$. We then transform each eigenvector by subtracting the $\phi_k^c$ that maximizes the coherence, obtaining a new $\phi_k\rightarrow \phi_k - \phi_k^c$ that is centered at 0. We also normalize negative and positive values such that the eigenvector loading becomes equally spaced on the interval $[-1,0]$ and $[0,1]$. All eigenvectors $\phi_k$ of the reversibilized $T_\text{ensemble}$ in Figs.\,\ref{fig:2},\ref{fig:3},\ref{fig:5},\ref{Suppl:2} are reported after performing these transformations. Finally, we use the notion of kinetic maps \cite{noe2015kinetic,noe2016commute} to convert dynamics in the space of these eigenvectors into equivalent distances in Euclidean space using this subsequent transformation,

\begin{align*}
    \psi_k = \sqrt{\frac{-\tau^*}{2\log(\lambda_k)}} \phi_k,
\end{align*}
where $\lambda_k$ is the eigenvalue corresponding to $\phi_k$ and $\tau^*$ is the transition time of the dynamics. We identify $q$ motor strategies by performing \emph{k-means} clustering on the $k$-dimensional space defined by the transformed eigenvectors $\psi_k, k \in [1,\lceil{\log_2 q}\rceil]$, weighted by the steady state invariant measure of each microstate. The \emph{k-means} clustering is performed using the \emph{scikit-learn} python package \cite{scikit-learn} over 1000 automatic repetitions with a \emph{k-means++} initialization. Coarse-graining at $q=2,4,7$ is reported in Fig.\,\ref{Suppl:2}. Further coarse-graining for a larger number of states is used in Fig.\,\ref{fig:4}.

\subsection*{Analyzing the kinematics encoded long-lived modes of $T_\text{ensemble}$}
We find correlations between the long-lived eigenvectors $\phi_k$ and interpretable kinematic variables in Fig.\,\ref{fig:2}(B,C,D). For this, we calculate the mean of each kinematic variable in the stacks of $K^*$ bouts that belong to a particular cluster $s_i,\, i \in \{1,\ldots,N^*\}$. Similarly for the inter-bout interval in Fig.\,\ref{Suppl:2}(H), we report the median value of the distribution of inter-bout intervals belonging to a particular cluster. For the coarse-grained strategies, we collect every trajectory of different fish executing one of the metastable strategies and bootstrap 100 times to calculate the cumulative distributions functions of the mean absolute change in heading and the mean speed with error bars in Figs.\,\ref{fig:2}(E),\ref{Suppl:2}(D). For the preference for egocentric direction in Fig.\,\ref{Suppl:2}(F), we assign each bout in a trajectory as left or right if the change in orientation from the start of one bout to the next is positive or negative respectively. The ratio of left or right movements in a trajectory is then calculated, and bootstrapped 100 times to report the mean value with error bars.

\subsection*{Simulations of symbolic sequences}
We use individual transition matrices of each fish at various coarse-graining scales $T_f^q$ to simulate the symbolic dynamics of the fish. This translates to learning $T_f^q$ for each fish from its individual symbolic sequence at a particular $q$ and then evolving the symbolic dynamics according to $T_f^q$ from its start state. We repeat this process a 100 times for each fish. We then calculate the mean sequence length in these simulations in each strategy at $q=2,4,7$ by bootstrapping across the 100 simulations for each fish.
(Fig.\,\ref{fig:2}(F),\ref{Suppl:3}).

\subsection*{Sampling lab space trajectories from symbolic sequences}
In order to simulate artificial trajectories from symbolic sequences, we sample velocity vectors (of the head position of the fish from the start of one bout to the start of the next) from the distribution of vectors $\rho^i_{\vec{v}}$ of a particular microstate $s_i,\, i\in\{1,\ldots,N^*\}$. For the symbolic sequences of real data, we simply sample $\vec{v}_i(t)\sim \rho^i_{\vec{v}}$ according to the distribution of velocities $\rho^i_{\vec{v}}$ within each visited discrete state $s_i(t)$ for increasing times $t$ (Fig.\,\ref{Suppl:8}A). To compare simulations with data in Fig.\,\ref{Suppl:7}, we utilize transition matrices $T$ with all the microstates $N^*$ at $\tau=1$ instead of $\tau=3$, which provides access to the fine-scale dynamics while retaining the information about the long-lived modes. We calculate this transition matrix $T$ using the ensemble of fish imaged in a $5\times 5\,\text{cm}^2$ arena in order to minimize mixing boundary effects from velocity vectors in different arenas. Finally, we simulate 1000 bout long symbolic sequences according to this transition matrix $T$, sampled from random initial conditions, and sample velocity vectors from the underlying distributions within each if the sampled microstates. Three example trajectories are shown in Fig.\,\ref{Suppl:8}B.

\paragraph{}
For the mean squared displacement (MSD) in Fig.\,\ref{Suppl:8}C, we report the MSD estimated from real trajectories (Real data in black from fish in a $5\times 5\,\text{cm}^2$ arena), artificial trajectories that effectively remove boundary conditions by sampling velocity vectors from the real symbolic sequence of the ensemble of fish (Real seq. in lime) and artificial trajectories obtained from simulated symbolic sequences (Sim seq. in magenta). In addition, we report the result of enforcing reflective boundary conditions on simulated symbolic sequences, limiting the arena size to $5\times 5\,\text{cm}^2$ (Refl. sims in red). 

\paragraph{}
For the long simulations from metastable strategies at the coarse-graining scale $q=4$, we  estimate an ensemble transition matrix restricted to the microstates that belong to the metastable strategy, using $\tau=1$, and sample from the inferred transition matrix 1000-bout long symbolic sequences 5000 times. Using the obtained symbolic sequences, we then sample from the underlying velocity vector distributions to get artificial lab space trajectories. The mean squared displacement calculations are reported in Fig.\,\ref{Suppl:8}D. For simulations from behavioral groups $\mathbb{G}=g$, we similarly build ensemble transition matrices $T^g$ at $\tau=1$ of fish belonging only to a specific group and then simulate 1000-bout long symbolic sequences  5000 times and sample velocity vectors as before.

\subsection*{Examining the preference for metastable strategies across conditions}
To investigate the preference for a strategy in a particular sensory context, we project the symbolic sequence of all fish in that condition along $\phi_1, \phi_2$. We then calculate a histogram of this projection with 50 bins along each axis, smoothed by a Gaussian kernel with a window size of 3 bins.

\subsection*{A distance metric to compare transition matrices}
Given a fish's individual transition matrix $T_q^f$ at a particular coarse graining scale $q$, we aim to compare dynamics of remaining in a particular metastable state or transitioning out of it, with that of another fish with transition matrix $T_q^{f'}$. We use the $L1$ norm (Manhattan distance) between every row of the transition matrices $T_q^f, T_q^{f'}$ to calculate these distances. With a distance value for each row of the two matrices $T_q^f$ and $T_q^{f'}$, we then calculate an average of these distance values to receive one value to compare fish to each other. Formally speaking, we have


\begin{align}\label{eq:distance}
    d(T_q^f, T_q^{f'}) = \frac{1}{q}\sum_i^{q}\sum_j^q|T_{ij}^f - T_{ij}^{f'}|,
\end{align}

which qualifies the criteria for being a metric.

\subsection*{Constant Shift Embedding for distance matrices}
Using the distance metric defined above, we are able to compare distances across 463 fish and build a distance matrix $D_q$ at a particular coarse-graining scale $q$ (datasets from \cite{marques2018structure}). We embed the calculated distance matrix between fish into a space where the Euclidean metric preserves the distances between fish using a Constant Shift Embedding (CSE)(\cite{roth2003optimal}). Unlike methods like Multidimensional Scaling (MDS) (\cite{kruskal1964nonmetric}) which minimize a cost function, CSE relies on solving an eigenvector problem, where each eigenvector for eigenvalues greater than 0 correspond to the dimensions of the Euclidean embedding. The eigenvalues of the CSE operation provide us with importance weights for each eigenvector. 

\subsection*{Logistic Regression to determine the correct coarse-graining scale to compare fish}
We attempt to classify each fish into its respective sensory context using only its behavior. We do so by comparing the test accuracy of a logistic regression at different levels of coarse-graining $q$, from 2 metastable strategies to a 1000. To do this, we first build individual transition matrices $T_q^f$ of all fish $f$ at a particular coarse-graining scale $q$ and transition time $\tau$. The coarse-graining $q$ is discovered in the same way as previously described from the eigenvectors of the ensemble transition matrix $T_{\text{ensemble}}$. We then estimate a distance matrix $D_q$ among all $T_q^f$, and perform CSE to embed it in a Euclidean space. We project the distance matrix to all the eigenvectors discovered by CSE with eigenvalues greater than 0.0. We then train logistic regressions (using the scikit-learn package \cite{scikit-learn}) over 100 shuffles of the data using a 80\%-20\% train-test split. Since each sensory context in the dataset from \cite{marques2018structure} contains different number of recordings, we shuffle recordings within each sensory context for every random seed and then sample 80\% of fish from each context and leave 20\% for testing. The logistic regression loss is weighted to account for this class imbalance. For each run of the logistic regression, we perform a grid search for the correct value of the $L2$ regularization parameter $\alpha$ using a 5-fold cross-validation (all of these steps are performed automatically by the scikit-learn GridSearchCV module). The model with highest validation accuracy is subsequently applied on the test data. We then compare the mean accuracy over the training and test sets (all weighted by the sample weight) over 100 shuffles of the data (Fig.\,\ref{fig:4}A). We also extract confusion matrices for each run of the logistic regression over a shuffle of the data and report the mean confusion matrix across the seeds (Fig.\,\ref{fig:4}B).

\subsection*{Hierchical multiplicative diffusive (HMD) clustering}
After selecting the correct coarse-graining scale $q=7$ based on the logistic regression, we work with the distance matrix $D_q$ obtained with $q=7$. We introduce a novel approach to perform a top-down clustering of the transition matrix space that takes into account the uncertainty in the estimate of the distances between data points. We do this by estimating an effective significance scale for each fish $\hat{\epsilon}_{f}$. This amounts to re-estimating transition matrices $\hat{T}_q^{f}$ from a 100 different simulations of symbolic sequences from $T_q^f$ and calculating the average distances between the $T_q^f$ and the re-estimated $\hat{T}_q^{f}$. Thus we have 

\begin{align*}
    \hat{\epsilon}_{f} = \langle d(\hat{T}_q^{f}, T_q^f) \rangle,
\end{align*}
where $d(\cdot,\cdot)$ is the distance defined in Eq.\,\ref{eq:distance}, and the expectation value is taken with respect to the 100 simulations. In Fig.\,\ref{Suppl:5}A, we provide the $\hat{\epsilon}_{f}$ for each fish. With this effective significance scale, we effectively rescale the distance among fish, and treat the top-down clustering as a search for metastability in a multiplicative diffusion process defined by the kernel 
\begin{align*}
    k(f_i,f_j) \propto \exp\left[D_{ij}/\sqrt{\hat{\epsilon}_i\hat{\epsilon}_j}\right].
\end{align*}
    
This operator rescales the distances between data points based on their uncertainty, such that points that have overlapping significance scales are effectively closer to each other than data points that do not overlap. For the first iteration of the hierarchical clustering, we leverage the first non-trivial eigenvector of this operator (the Fiedler vector) that partitions the data points into two (\cite{froyland2003detecting}) by finding the location of an effective barrier height in the diffusion dynamics. This is effectively equivalent to performing spectral clustering into two clusters using a 1-D diffusion map of data. To obtain a fuzzy cluster assignment, we discover this clustering by fuzzy c-means clustering (\emph{skfuzzy} package \cite{warner2019}). We repeat this process iteratively: At the second iteration, we first create two multiplicative diffusion operators using only the data in each cluster from the first iteration. We then quantify the effective barrier height of these diffusive processes by measuring the level of metastability of the diffusive dynamics within the cluster. Effectively, this is calculated in the same way as the dynamics, by maximizing the coherence along the first non-trivial eigenvector of the diffusion operator. Then we split the cluster that maximizes the metastability. Because we discover a fuzzy clustering assignment, we essentially receive posterior distributions for each iteration. However, since these posteriors are limited the cluster that was split, we extend them other data points by estimating distances between the split cluster center and the unseen data points and converting them to probabilities. This is handled automatically by the skfuzzy package.

\subsection*{Estimating kinematic parameters for phenotypic groups $\mathbb{G}$}
We convert the posterior $P(\mathbb{G}=g|f)$ to $P(f|\mathbb{G}=g)$ using Bayes rule and calculate subsequent kinematic parameters as expectation values. For example, for the case of Fig.\,\ref{fig:7}, we calculate the percentage time spent in eye convergence $ec$ in a sensory context $c$ as -

\begin{align*}
    \mathbb{E}[ec(f|\mathbb{G}=g, c)] = \sum_{f_i} ec(f|c)P(f|\mathbb{G}=g,c)
\end{align*}

We estimate errorbars by utilizing importance sampling, where we repeatedly estimating the mean of $ec(f|\mathbb{G}=g, c)$ by drawing $f$ from $P(f|\mathbb{G}=g,c)$.

\label{s:methods}

\section*{Author Contributions}
G.S., M.V., A.C.C and C.W. conceived and designed the project. G.S. and A.C.C performed the analysis. G.S, J.C.M and M.B.O curated and preprocessed the datasets. G.S., A.C.C and C.W. wrote the paper. All authors reviewed the results and approved the final version of the manuscript.

\section*{Acknowledgements}
We would like to thank all the members of the Wyart lab for extensive discussions, especially Dr. Martin Carbo-Tano, Dr. Elias T. Lunsford, and Mahalakshmi Dhanasekar for their insight in fish neuroanatomy and behavior. G.S was supported by the European Union’s Horizon 2020 Research and Innovation program under the Marie Skłodowska-Curie Grant No. 813457 awarded to C.W and managed by Dr. Joana Guedes (https://zenith-etn.com) at the Paris Brain Institute (Institut du Cerveau). G.S, C.W and A.C.C were also supported by Fondation Bettencourt-Schueller (FBS-don-0031), the European Research Council (ERC Consolidator ERC-CoG-101002870), the team grant Fondation pour la Recherche Médicale (FRM-EQU202003010612), the Agence Nationale pour la Recherche (ANR) LOCOCONNECT, RocSMAP, ASCENTS, MOTOMYO. G.S and C.W also acknowledge support in part by the grant NSF PHY-1748958 of the Kavli Institute for Theoretical Physics (KITP) and the Gordon and Betty Moore Foundation Grant No. 2919.02. A.C.C was supported by LabEx ENS-ICFP: ANR-10-LABX-0010/ANR-10-IDEX-0001-02 PSL* and M.V by the NIH Grant No. 1RF1NS128865-01. J.C.M received the support of a fellowship from “la Caixa” Foundation (ID100010434, LCF/BQ/PR21/11840005) and from the Portuguese Foundation for Science and Technology (EXPL/MED-NEU/0957/2021). M.B.O. was supported by grants from the Volkswagen Stiftung “Life?” Initiative (A126151) and ERC (NEUROFISH 773012).

\section*{Bibliography}

\bibliographystyle{bxv_abbrvnat}
\bibliography{refs.bib}

\begin{figure*}[ht]
\centering
\includegraphics[width = 1.\textwidth]{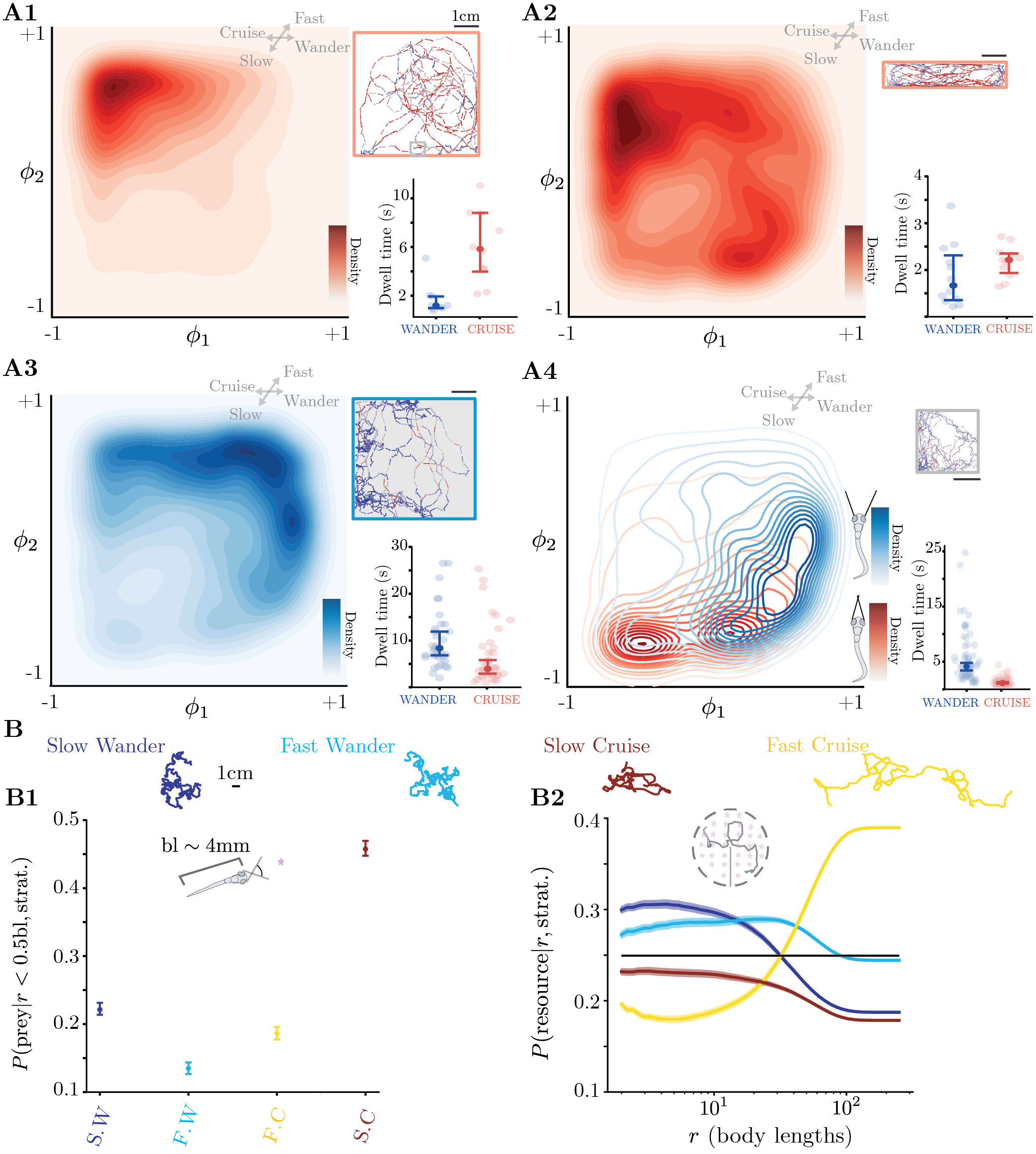}
\end{figure*}

\begin{figure*}[ht]
\centering
\caption{\textbf{Functional role of motor strategies across sensory contexts.} \textbf{(A)} Probability density of visiting different bout sequences along $\phi_1-\phi_2$ (left). As insets, an example fish trajectory color coded by $\phi_1$(top right) and the median dwell time in each motor strategy (bottom right), where the error bars represent 95\% confidence intervals bootstrapped across the fish belonging to a sensory context (plotted individually in the background). \textbf{(A1)} In light ($1000\,\text{lm/m}^2$, $5\times 5\,\text{cm}^2$ arena, 10 fish), fish mostly perform fast cruising strategies. \textbf{(A2)} In a smaller arena ($1000\,\text{lm/m}^2$, $1\times 5\,\text{cm}^2$  arena,12 fish), we observe a higher usage of wandering strategies, which is particularly enhanced at the short ends where fish are force to quickly reorient (the width of the arena is only about twice the body length of the fish, inset). This results in a reduction of the time spend cruising when compared with the $5\times 5\,\text{cm}^2$ arenas in panel A1. \textbf{(A3)} Fish in the dark ($5\times 5\,\text{cm}^2$, 37 fish) show a shift towards fast wandering behaviors. We combine data from two different conditions: the ``Dark'' condition, in which fish are simply freely-swimming in the dark, and also the first 30 minutes of the ``Dark Transitions'' condition, in which fish are also freely-swimming in the dark before being exposed to light of different intensities (see Table.\,\ref{Suppl:Table}). 
\textbf{(A4)} In a prey capture assay with $\approx 50$ paramecia in the arena ($1000\,\text{lm/m}^2$, $2.5\times 2.5\,\text{cm}^2$  arena, 65 fish), fish tend to engage in slow cruising behaviors during eye convergence events, while in the inter-hunt period they mostly perform wandering behaviors. Notably, there is a significant shift from free exploration in the light (A1), with the near absence of fast cruising behaviors even in the inter-hunt period.
\textbf{(b)} Probability of gathering resources in two distinct regimes: one at short length scales in which the fish has its eyes converged and tries to capture prey within its field of view ($60^\circ$ aperture), and another at large length scales in which we assume the fish can sense resources all around its body (see Methods). We compare the $q=4$ motor strategies, which correspond to splitting the Cruising and Wandering strategies into their slow and fast variants (Fig.\,\ref{Suppl:2}D). We simulate the dynamics in each motor strategies by inferring transition matrices using only microstates belonging to each of the motor strategies and data from all fish. We then simulate trajectories with a length of 1000 bouts and repeat the process 5000 times starting from random initial conditions. 
(\textbf{B1}) At the scale of less than half the body length, in which the fish begins prey pursuit with its eyes converged (\cite{mcelligott2005prey}), slow cruising is the most successful strategy at acquiring prey.
(\textbf{B2}) At length scales of larger than twice the body length and while searching all around the body, wandering strategies are effective for short to meso-length scale searching, while fast cruising is effective at large scale dispersal. Black line represents the behavior of an average fish, which is equally efficient at all length scales.
}
\label{fig:3}
\end{figure*}

\begin{figure*}[ht]
\centering
\includegraphics[width = 1.\textwidth]{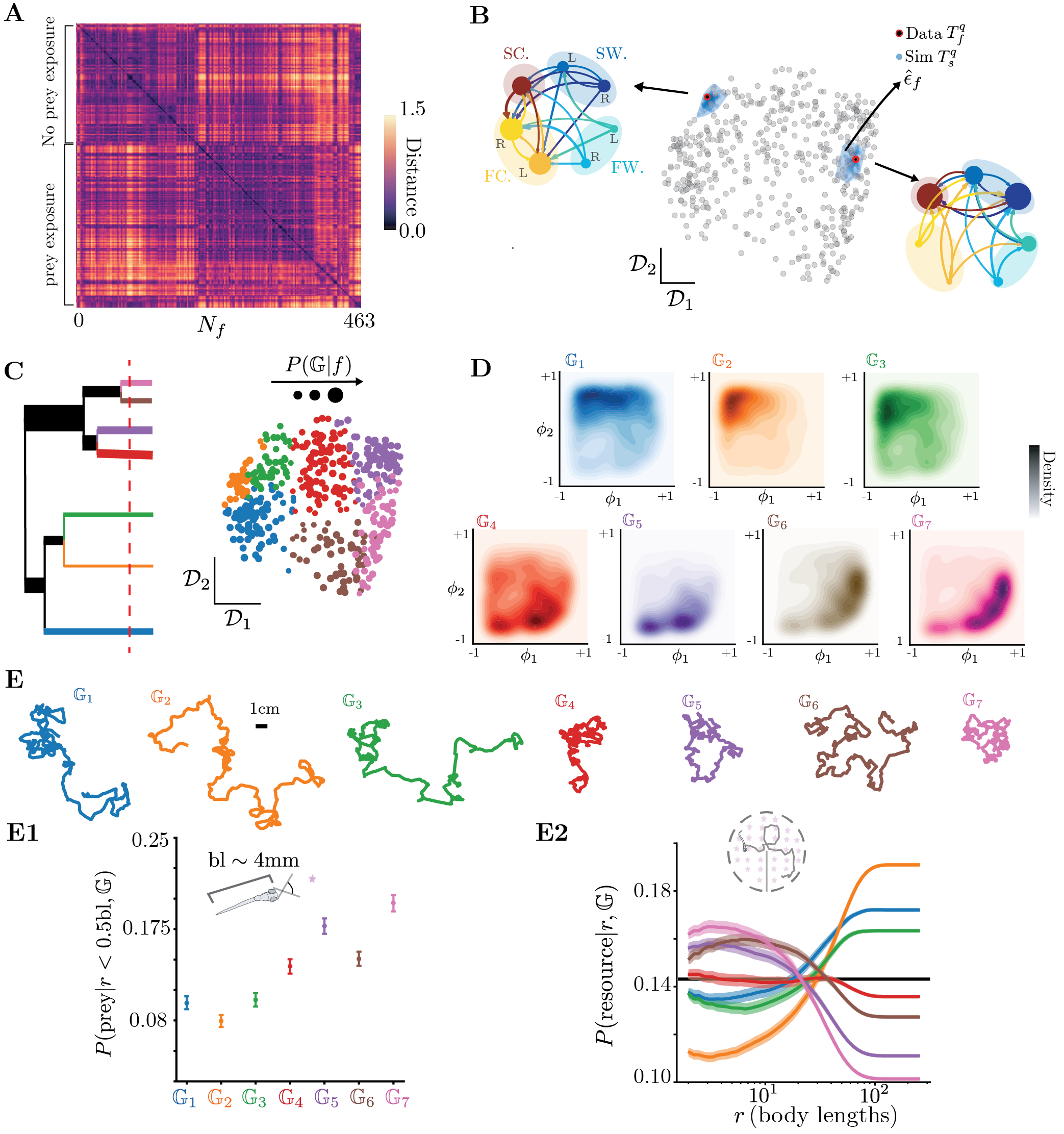}
\end{figure*}

\begin{figure*}[ht]
\centering
\caption{\textbf{Revealing phenotypic groups by clustering fish based on their behavioral dynamics at the experimental timescale} 
\textbf{(A)} 
Pairwise distance matrix among the transition matrices of each fish $T_q^f$ at $q=7\,\text{states}$ (see Methods). We compare individual fish to each other based purely on their behavioral dynamics (see Methods), finding similaries among fish at several scales. The broad block diagonal structure of the matrix indicates that sub-groups of fish behave similarly, with prey exposure seeming as the broadest determinant of this variability. However, the distance matrix is also highly heterogeneous, suggesting structure beyond the sensory context.
\textbf{(B)}  We use Constant Shift Embedding (CSE) \cite{roth2003optimal} to embed the phenotypic space into an Euclidean space that preserves the pairwise distances among fish (see Methods). We visualize each individual fish (gray) along the first two dimensions of this \emph{transition matrix space}. We also provide two example transition matrices of fish from different parts of the space. To reveal the multiscale structure to the space of behavioral phenotypes, we estimate whether pairs of fish behave significantly different from each other within the finite time of the experiment (see Methods). We re-estimate transition matrices from finite simulations of each fish (with the same length as the original recordings), and estimate the distance between re-estimated transition matrices and the original transition matrix (blue points surrounding the example fish). Averaging these distances sets an effective scale $\hat{\epsilon}_f$ (blue area around example fish) within which neighboring fish are indistinguishable from each other (see Methods).
\textbf{(C-left)} We reveal the structure in the phenotypic space through a top-down fuzzy subdivision of a multiplicative diffusion process (HMD clustering), in which distances are rescaled by $\hat{\epsilon}_f$ (see Methods). At each iteration, we identify the group of fish that is most separable (see Methods), and subdivide it. In this way, the ordering of sub-divisions is indicative of the relative scale separation among fish from different groups.  We illustrate our clustering approach as a tree diagram. We stop the clustering after 6 levels (7 clusters), the point beyond which the effective distances between fish in clusters stops growing (see Fig.\ref{Suppl:5}B). The widths of the branches of the tree are proportional to the number of fish in each cluster. \textbf{(C-right)} We color-code each individual fish by their most likely phenotypic group $\mathbb{G}_i$, where $i\in\{1,\ldots,7\}$. The size of the dot indicates the posterior probability $P(\mathbb{G}^{L=6}|f)$.
(\textbf{D}) The clustering reveals significantly different phenotypic groups $\mathbb{G}_i$, which nonetheless share common behavioral structures according to the hierarchical subdivision. Groups $\mathbb{G}_1$, $\mathbb{G}_2$ and $\mathbb{G}_3$ all exhibit a bias towards fast cruising, but vary in how much they also use fast wandering ($\mathbb{G}_1$) and slow cruising and wandering ($\mathbb{G}_3$) strategies. Groups $\mathbb{G}_4$, $\mathbb{G}_5$ share a bias towards slow cruising and wandering behaviors, with $\mathbb{G}_4$ exhibiting a higher proportion of faster cruising. Finally, groups $\mathbb{G}_6$, $\mathbb{G}_7$ are both biased towards fast wandering strategies, with $\mathbb{G}_7$ exhibiting more slow cruising behaviors.
\textbf{(E)} Probability of gathering resources in two distinct regimes, as in Fig.\,\ref{fig:3}(B) (see Methods). We compare the $7$ phenotypic groups, simulating the dynamics in each group by inferring transition matrices across all fish that belong to a particular phenotypic group. We simulate trajectories with a length of 1000 bouts and repeat the process 5000 times starting from random initial conditions. 
\textbf{(E1)} Probability of capturing resources uniformly distributed in a distance shorter than half a body length and within a cone of $60^\circ$ ahead of the fish (see Methods). Groups $\mathbb{G}_5$ and $\mathbb{G}_7$ are the most effective at pursuing and capturing prey, reflecting their higher biases towards slow cruising and wandering strategies.
\textbf{(E2)} Probability of finding resources uniformly distributed in a distance shorter than $r$ (see Methods). Groups $\mathbb{G}_{5,6,7}$ are effective for short to meso-length scale searching, while $\mathbb{G}_1$ and $\mathbb{G}_2$ are most effective at large scale dispersal. Group $\mathbb{G}_4$ is approximately equally efficient across a broad range of length scales. Black line represents all groups being equally efficient.
} 
\label{fig:5}
\end{figure*}

%% file: 02_Article_Supplementary.tex
\onecolumn 
\fancyhead{} 
\renewcommand{\floatpagefraction}{0.1}
\lfoot[\bSupInf]{\dAuthor}
\rfoot[\dAuthor]{\cSupInf}
\newpage

\captionsetup*{format=smallformat} 
\setcounter{figure}{0} 
\setcounter{equation}{0} 
\makeatletter 
\renewcommand{\thefigure}{S\@arabic\c@figure} 
\makeatother
\def\theequation{S\arabic{equation}}

\newpage
\section*{Supplementary Information}

\begin{table}[h!]
\begin{center}
\caption{Different Sensory conditions from \cite{marques2018structure}}
\label{Suppl:Table}
\begin{tabular}{|p{6cm}||p{1cm}|p{2.5cm}|}
 \hline
 Sensory Condition & Number of fish  & Arena Size (W cm x H cm x D cm)  \\
 \hline
 Light & 10  & 5 x 5 x 0.3\\
 Light &  12 & 1 x 5 x 0.8\\
 Expanding Spot &61 &5 x 5 x 0.3\\
 Dark Transition & 27 & 5 x 5 x 0.3 \\
 Phototaxis & 30 & 5 x 5 x 0.3\\
 Forward OMR & 12 & 1 x 5 x 0.8\\
 Directional OMR & 30 & 5 x 5 x 0.3\\
 Dark & 10 & 5 x 5 x 0.3\\
 High Lux Light/Dark Transitions & 22 & 5 x 5 x 0.3\\
 Prey Capture Paramecia Naive & 69 & 2.5 x 2.5 x 0.3\\
 Prey Capture Paramecia Raised With & 98 & 2.5 x 2.5 x 0.3\\
 Prey Capture Rotifer Naive & 15 & 2.5 x 2.5 x 0.3\\
 Prey Capture Rotifer Raised With & 31 & 2.5 x 2.5 x 0.3\\
 Light Rotifer Raised With & 60 & 2.5 x 2.5 x 0.3\\
 \hline
\end{tabular}
\end{center}
\end{table}

\begin{figure}[h!]
\centering
\includegraphics[width = 1.\textwidth]{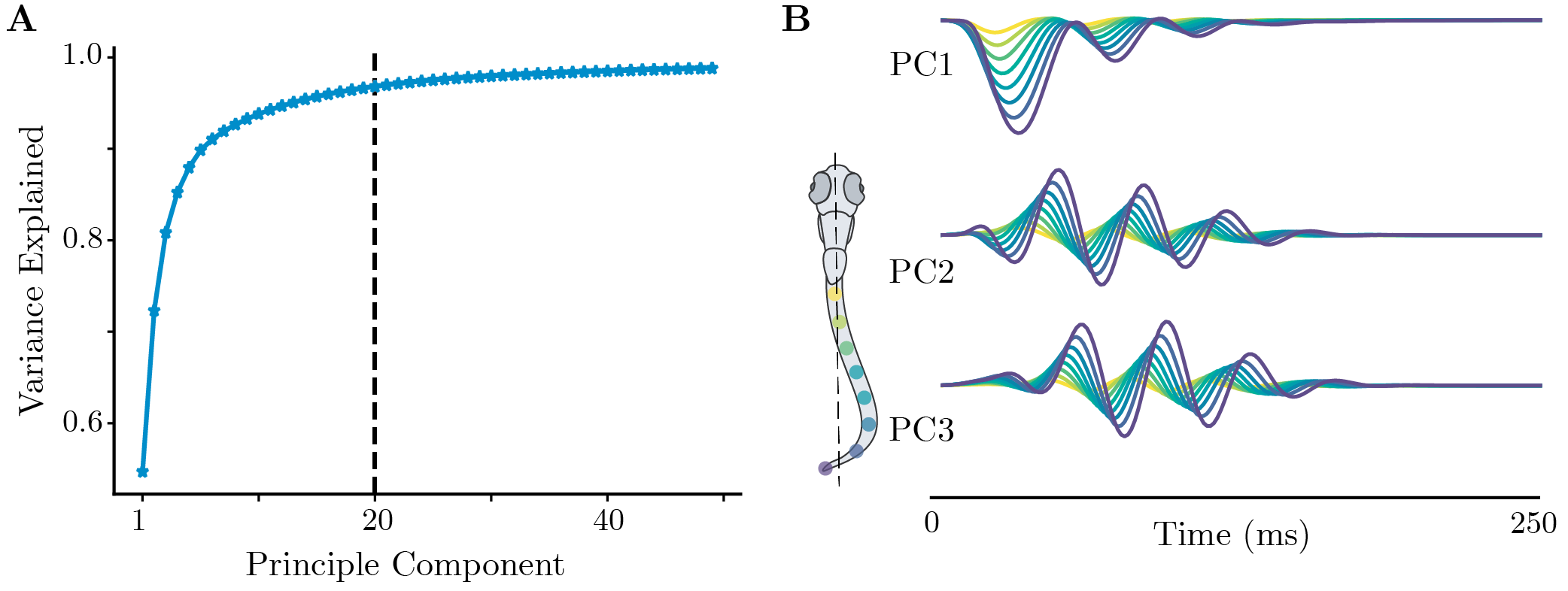}
\caption{\textbf{Principle component analysis (PCA) of bout space.} \textbf{(A)} All bouts are represented by the cumulative tail angle of 8 points on the tail for $250\,\text{ms}$ from the start of the bout. For the dataset from \cite{marques2018structure}, a single bout is then composed of $175\,\text{frames}\times 8$ (at a sampling frame rate of 700Hz). We randomly sample 25 recordings across all sensory conditions ($\approx 50,000$ bouts per sampling), estimate a covariance matrix and obtain its eigenvalues and eigenvectors. We then estimate the mean covariance matrix and the means of the eigenvalues and eigenvectors across all such possible samples of the data. We find that the first 20 eigenvectors capture $>$95\% of the variance. We also calculate 95\% errorbars on the estimate of the variance explained across multiple resampling, but these are too minuscule to notice. \textbf{(B)} To showcase what the PCA eigenvectors represent, we plot the first three principle components. The first component PC1 mostly captures turning biases (leftward vs rightward turns), the second and third components, PC2 and PC3, correspond to forward waves. Finer scale details of rarer bouts are captured by the remaining modes.} 
\label{Suppl:1}
\end{figure}

\begin{figure}[ht!]
\centering
\includegraphics[width = 1.\textwidth]{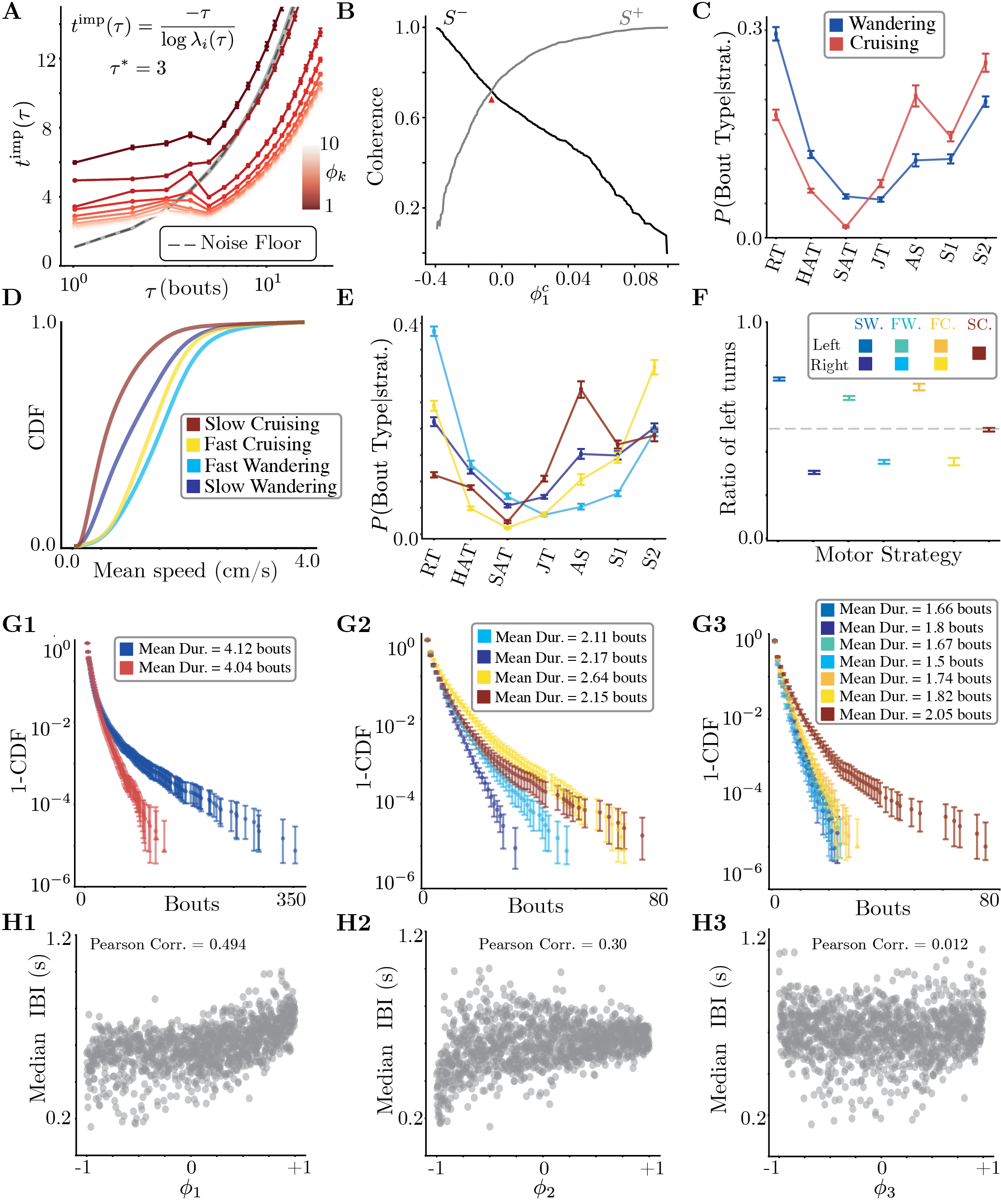}
\end{figure}

\begin{figure}[h!]
\centering
\caption{\textbf{Details of extracting and analyzing the long lived modes from the transition matrix $T_{\text{ensemble}}$} \textbf{(A)} Implied timescales $t_\text{imp}(\tau)$ as a function of the transition time $\tau$. We estimate the implied timescales by randomly sampling 7500 bouts from each of the 14 conditions over 100 random seeds. Errorbars are bootstrapped 95\% estimates of the implied timescales. For each value of $\tau$, we estimate a transition matrix by counting the number of times the dynamics goes from state $s_i$ to $s_j$ after $\tau$ bouts. When $\tau$ is too large, the dynamics starts to effectively mix, meaning that samples that are too disparate in time are practically independent from each other. In this limit,  there should only be one surviving non-zero eigenvalue of $T_{ij}$, corresponding to the steady-state distribution. However, the finite length of the recordings induces a non-zero noise floor, which we gauge by estimating the largest eigenvalue from a transition matrix built from shuffled symbolic sequences (gray line). At smaller $\tau$, a large fraction of eigenvalues are significant, and we choose an intermediate $\tau^*=3\,\text{bouts}$ to isolate the slow modes from the bulk of the spectrum. 
\textbf{(B)} Coherence of each metastable strategy a function of $\phi_1^c$. We scan along $\phi_1$ and define $S^-$ and $S^+$ as collections of microstates that have $\phi_1$ smaller or larger than a threshold $\phi_1^c$, respectively. We then estimate the fraction of probability that remains within both sets as we vary $\phi_1^c$, which we call coherence (see Methods for details, \cite{froyland2005statistically}). We identify the longest lived motor strategies by identifying the value of $\phi_1^c$ that maximizes the coherence of both motor strategy.
\textbf{(C)} Probability of observing different bout types in each of the coarse-grained strategies: Cruising and Wandering. We find a bias towards turning behaviors in Wandering (\textbf{R}outine \textbf{T}urn, \textbf{H}igh \textbf{A}ngle \textbf{T}urn, \textbf{S}pot \textbf{A}voidance \textbf{T}urn), while Cruising behaviors are dominated by forward movements (\textbf{A}pproach \textbf{S}wim, \textbf{S}low \textbf{1}, \textbf{S}low \textbf{2}). Notice also that despite these broad differences, there is also a non-negligible fraction of routine turns (RT) during Cruising, as well as a large fraction of S2 in Wandering. 
\textbf{(D)} Cumulative Distribution Function (CDF) of the mean instantaneous speed in a bout sequence in the four long-lived motor strategies. We discover finer scale states by using faster decaying eigenvectors: here, we use $\phi_1$ and $\phi_2$ to reveal 4 motor strategies, which correspond to Slow and Fast variations of Cruising and Wandering (see Methods for details). 
\textbf{(E)} Probability of observing different bout types in each of the 4 motor strategies. 
\textbf{(F)} We further identify faster timescale motor strategies using the first three eigenvectors $\phi_k, k\in{1,2,3}$ (see Methods for details). We find left and right direction bias variations of the slow wandering, fast wandering and fast cruising states. In contrast, the slow cruising does not exhibit persistent left/right biases. 
\textbf{(G1)} Complementary cumulative distribution function (1 - CDF) of the time spent in the Cruising and Wandering states across all fish. We find a broad distribution of timescales, with a comparatively small average sequence length of $4.12 \,\text{bouts} \Rightarrow 3.35\,\text{s} \pm(3.32,3.39)\,\text{s}$ in Wandering and  $4.04 \,\text{bouts}\Rightarrow  2.67\,\text{s} \pm(2.64, 2.7)\,\text{s}$ in Cruising. Errorbars represent bootstrapped 95\% confidence intervals across fish. 
\textbf{(G2)} Complementary cumulative distribution function (CCDF) of the time spent in the slow and fast variation of Cruising and Wandering states across all fish. As expected, these states capture faster timescale properties of the bout sequence dynamics, with slow cruising lasting on average $1.04\,\text{s} \pm(1.02,1.06)\,\text{s}$, fast cruising $1.58\,\text{s}\pm(1.56, 1.60)\,\text{s}$, slow wandering $1.34\,\text{s}\pm(1.32,1.36)\,\text{s}$ and fast wandering $1.27\,\text{s}\pm(1.26,1.29)\,\text{s}$. Errorbars represent 95\% confidence intervals measure across fish.
\textbf{(G3)}  Complementary cumulative distribution function (CCDF) of the time spent in the left/right variation of the Slow/Fast strategies. Errorbars represent 95\% confidence intervals measure across fish. \textbf{(H1)} Median inter-bout interval in the bout sequences corresponding to a given microstate, plotted against $\phi_1$. We find that the inter-bout interval is correlated with $\phi_1$.
\textbf{(H2)} Median inter-bout interval in the bout sequences corresponding to a given microstate, plotted against $\phi_2$. We find that slow microstates have lower inter-bout intervals.
\textbf{(H3)} Median inter-bout interval in a microstate plotted against $\phi_3$. There is almost no correlation between the inter-bout interval and the egocentric direction preference.} 
\label{Suppl:2}
\end{figure}

\begin{figure}[h!]
\centering
\includegraphics[width = 1.\textwidth]{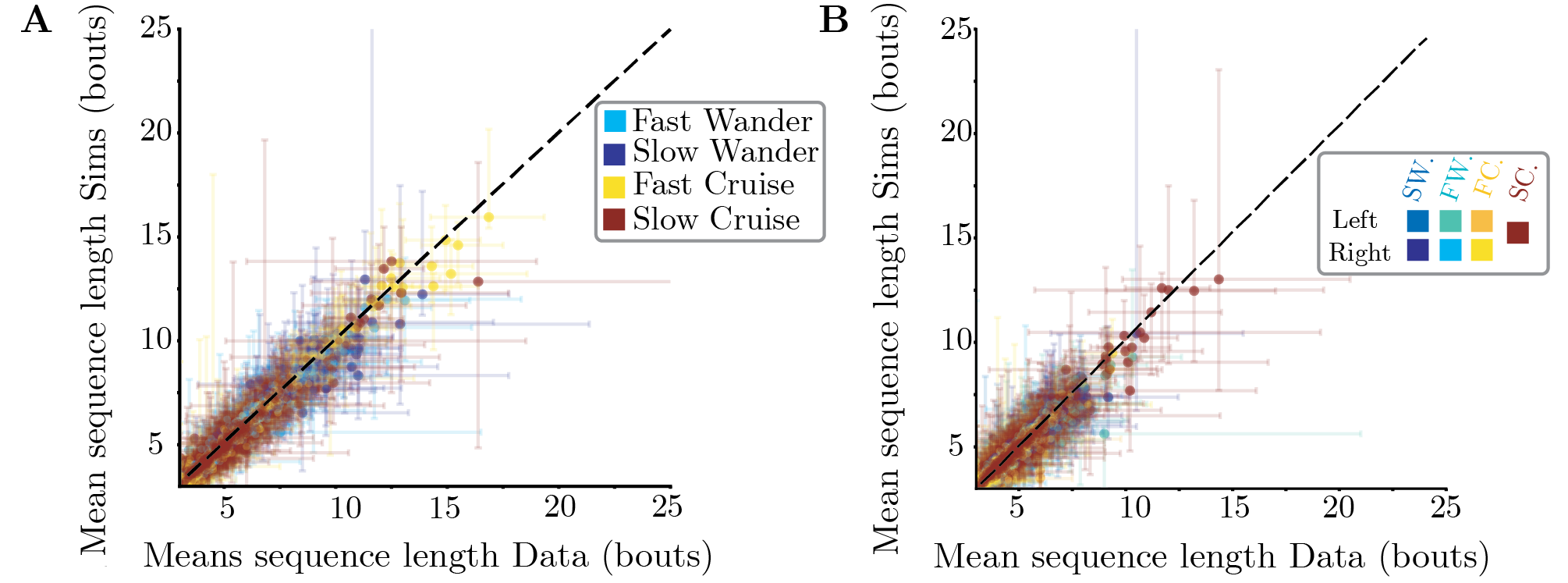}
\caption{\textbf{Simulations for individual fish at different coarse-graining scales} \textbf{(A)} Mean sequence length for data and simulations at the level of four metastable states. \textbf{(B)} Mean sequence length for data and simulations at the level of seven metastable states. We get a good match between the data and the simulations even at the scale of these shorter timescale strategies. Errorbars represent bootstrapped 95\% confidence intervals on the mean sequence length. To obtain error bars from the simulations, we collect behavioral events from 100 independent simulations, and bootstrap by sampling a number of events equal to the one observed in the real data (see Methods).} 
\label{Suppl:3}
\end{figure}

\begin{figure}[h!]
\centering
\includegraphics[width = 1.\textwidth]{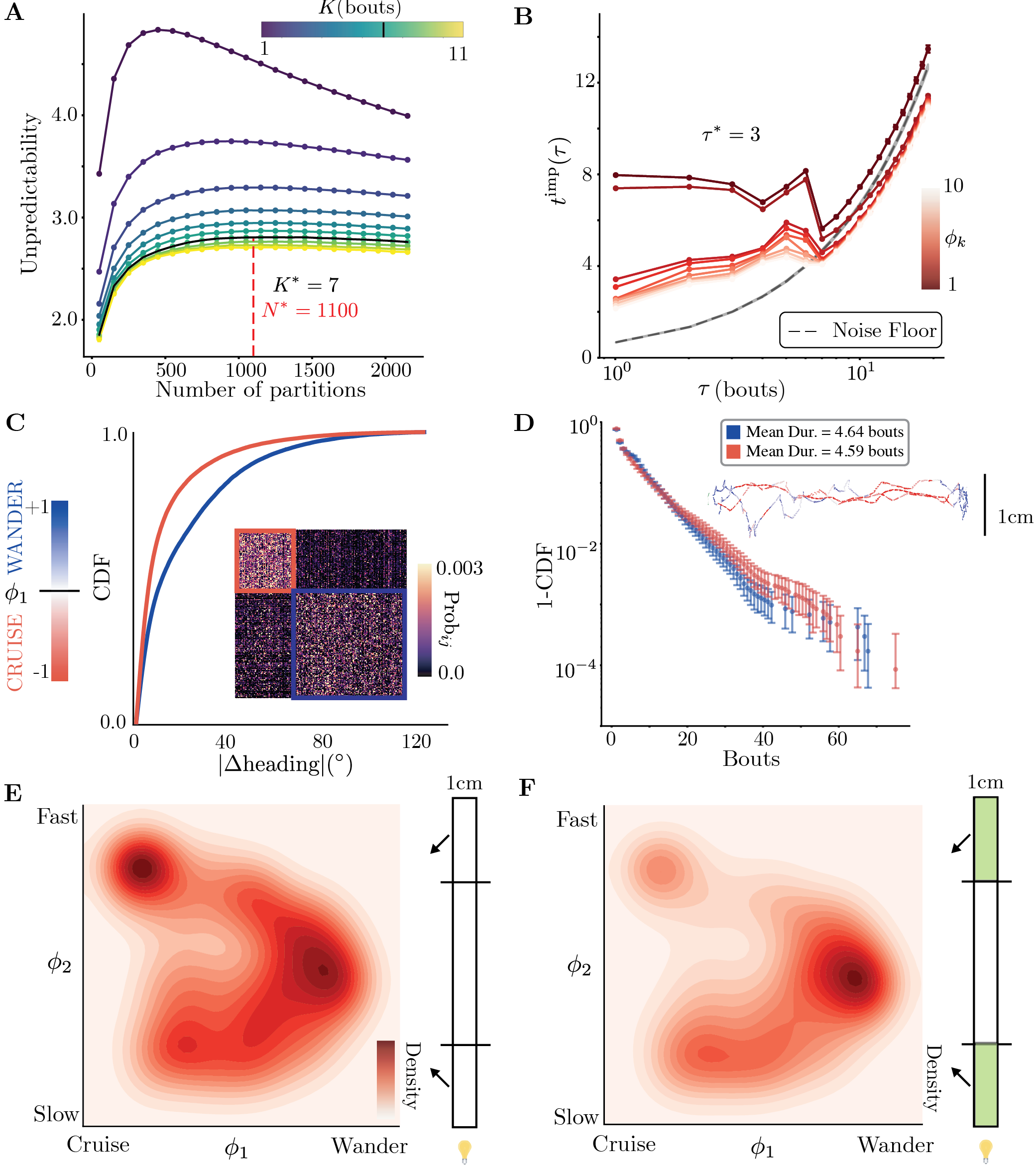} 
\end{figure}

\begin{figure}[h!]
\centering
\caption{\textbf{Discovering the long lived modes of behavior in the dataset from \cite{reddy2022lexical}}. We note that this dataset was collected at a lab different from the one that collected the data used in the main text \cite{marques2018structure}, and the bouts were tracked using a different tracking algorithm (Zebrazoom \cite{mirat2013zebrazoom}).  Here, the fish are swimming in a $1\,\text{cm}\times 14\,\text{cm}$ arena and some fish were subject to aversive acidic pH at the arena edges. We take all recordings where the fish swim for longer than 250 bouts, which gives 123 recordings without pH and 93 recordings with pH (see Methods).  
\textbf{(A)} Unpredictability as a function of the number of stacked bouts $K$ and the number of partitions. We measure unpredictability as in the main text: we equally sample 40,000 bouts from the two different sensory condition (with and without acidic pH), and estimate the entropy rate of the Markov chain constructed by partitioning the space defined by sequences of $K$ bouts into $N$ microstates. We find that in this dataset $K^*=7\,\text{bouts}$ minimizes the entropy rate, and $N^*=1100$ is the maximum number of partitions beyond which finite-size effects results in an underestimation of the entropy rate. Notice that, compared to Fig.\,\ref{fig:1}, we find a larger $K^*$, reflecting longer timescale correlations among the tracked bouts. Errorbars represent bootstrapped 95\% confidence intervals over 50 resamples of the data. 
\textbf{(B)} Implied timescales of the transition matrix at $K^*=7, N^*=1100$. We choose $\tau^*=3$ as before. 40,000 microstates are sampled from each condition. Errorbars represent bootstrapped 95\% confidence intervals over 20 resamplings of the symbolic sequences. 
\textbf{(C)} As in the main results, the first non-trivial eigenvector $\phi_1$ of the ensemble transition matrix coarse-grains the dynamics into Cruising-Wandering strategies. We plot the cumulative distribution of the mean absolute change in heading in each motor strategy. (inset) The coarse-graining results in a block-diagonal structure in the transition matrix. 
\textbf{(D)} Complementary cumulative distribution functions (1-CDF) of the sequence length in either Cruising or Wandering strategies with one example trajectory of 200 bouts color coded by $\phi_1$.
\textbf{(E)} Kernel density estimates along $\phi_1, \phi_2$ for the non-acidic control recordings, sampled only in regions where pH would have spread. Fish have a distributed preference for both cruising and wandering in these regions in the recordings.
\textbf{(F)} Kernel density estimates along $\phi_1, \phi_2$ for the acidic pH recordings, sampled only in regions where pH would have spread. Fish have a specific preference for wandering in these regions, pointing to wandering as a long lived response to the pH.} 
\label{Suppl:4}
\end{figure}

\begin{figure}[h!]
\centering
\includegraphics[width = 1.\textwidth]{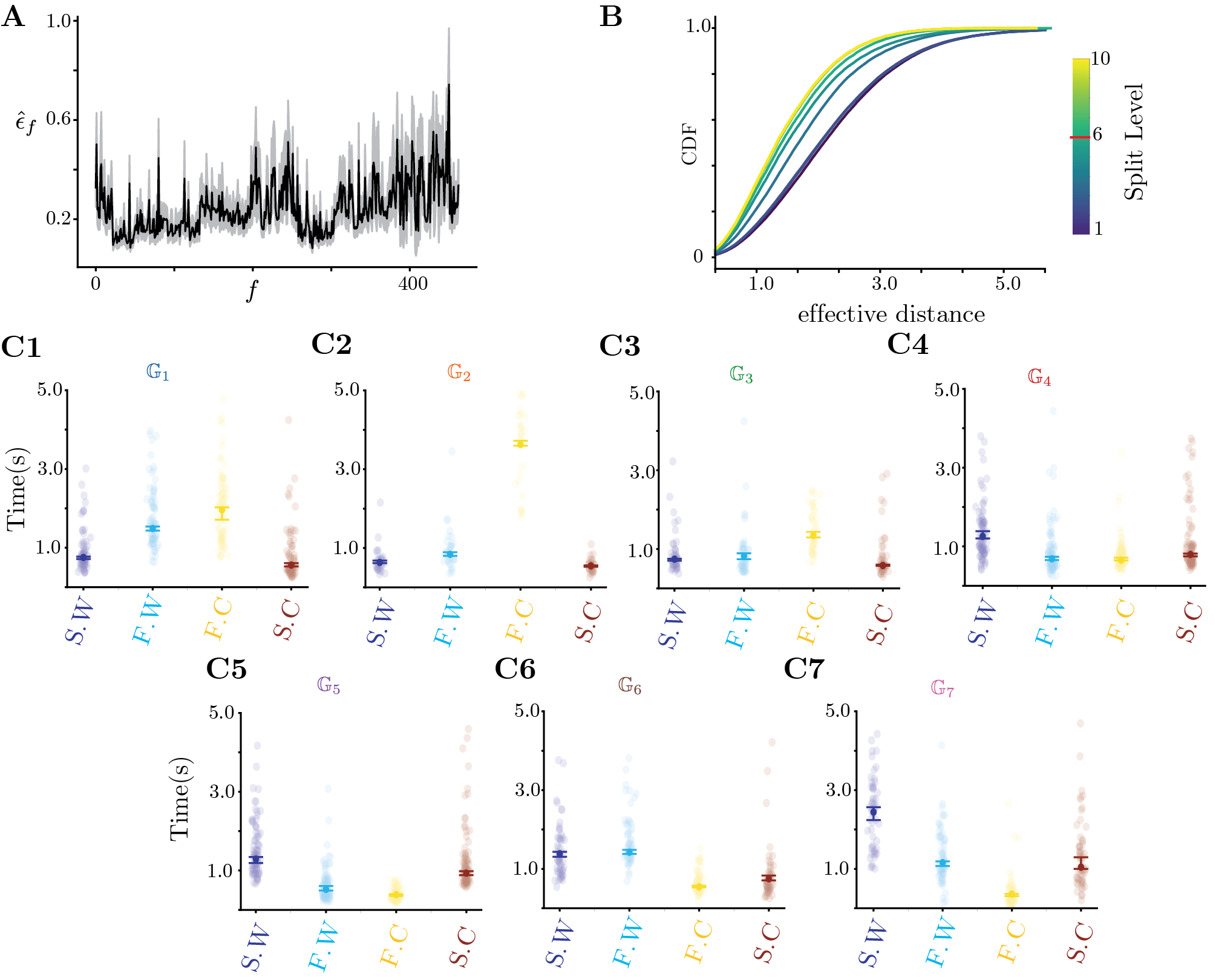}
\caption{\textbf{Details on the Hierarchical multiplicative diffusive clustering and the discovered phenotypic groups $\mathbb{G}$}  \textbf{(A)} The effective scale separation $\hat{\epsilon}_{f_i}$ for each fish at $q=7$ (see Methods). Light grey represent 95\% bootstrapped confidence intervals. \textbf{(B)} Cumulative distributions of effective distances within a cluster at different iterations $L$ of the hierarchical clustering. The distributions change very little after $L=6$, setting that as the cut-off point for the clustering. \textbf{(C)} Each panel shows the median dwell time across fish in either fast wandering, slow wandering, fast cruising or slow cruising for one of the four behavioral groups $\mathbb{G}$. We calculate the mean in each metastable strategy within a fish and take medians across fish. Each point represents the mean dwell time of a single fish, errorbars represent bootstrapped 95\%confidence intervals on the median across fish.
\textbf{(C1)} Median dwell times in $\mathbb{G}_1$. Slow Wandering: $0.75\,(0.71, 0.77)\,\text{s}$, Fast Wandering: $1.49\,(1.44, 1.59)\,\text{s}$, Fast Cruising: $1.96\,(1.71, 2.03)\,\text{s}$, Slow Cruising: $0.56\,(0.53, 0.61)\,\text{s}$ \textbf{(C2)} Median dwell times in $\mathbb{G}_2$. Slow Wandering: $0.65\,(0.62, 0.68)\,\text{s}$, Fast Wandering: $0.84\,(0.8, 0.89)\,\text{s}$, Fast Cruising: $3.63\,(3.59, 3.72)\,\text{s}$, Slow Cruising: $0.55\,(0.52, 0.56)\,\text{s}$
\textbf{(C3)} Median dwell times in $\mathbb{G}_3$. Slow Wandering: $0.74\,(0.71, 0.76)\,\text{s}$, Fast Wandering: $0.82\,(0.74, 0.89)\,\text{s}$, Fast Cruising: $1.35\,(1.3, 1.44)\,\text{s}$, Slow Cruising: $0.58\,(0.57, 0.6)\,\text{s}$
\textbf{(C4)} Median dwell times in $\mathbb{G}_4$. Slow Wandering: $1.25\,(1.2, 1.38)\,\text{s}$, Fast Wandering: $0.69\,(0.65, 0.74)\,\text{s}$, Fast Cruising: $0.65\,(0.64, 0.71)\,\text{s}$, Slow Cruising: $0.8\,(0.75, 0.82)\,\text{s}$
\textbf{(C5)} Median dwell times in $\mathbb{G}_5$. Slow Wandering: $1.3\,(1.19, 1.34)\,\text{s}$, Fast Wandering: $0.52\,(0.49, 0.61)\,\text{s}$, Fast Cruising: $0.39\,(0.35, 0.39)\,\text{s}$, Slow Cruising: $0.93\,(0.87, 0.99)\,\text{s}$
\textbf{(C6)} Median dwell times in $\mathbb{G}_6$. Slow Wandering: $1.4\,(1.3, 1.43)\,\text{s}$, Fast Wandering: $1.42\,(1.38, 1.48)\,\text{s}$, Fast Cruising: $0.54\,(0.53, 0.57)\,\text{s}$, Slow Cruising: $0.75\,(0.71, 0.85)\,\text{s}$
\textbf{(C7)} Median dwell times in $\mathbb{G}_7$. Slow Wandering: $2.44\,(2.19, 2.57)\,\text{s}$, Fast Wandering: $1.16\,(1.07, 1.19)\,\text{s}$, Fast Cruising: $0.36\,(0.31, 0.37)\,\text{s}$, Slow Cruising: $1.04\,(1.0, 1.29)\,\text{s}$
} 
\label{Suppl:5}
\end{figure}

\begin{figure}[h!]
\centering
\includegraphics[width = 1.\textwidth]{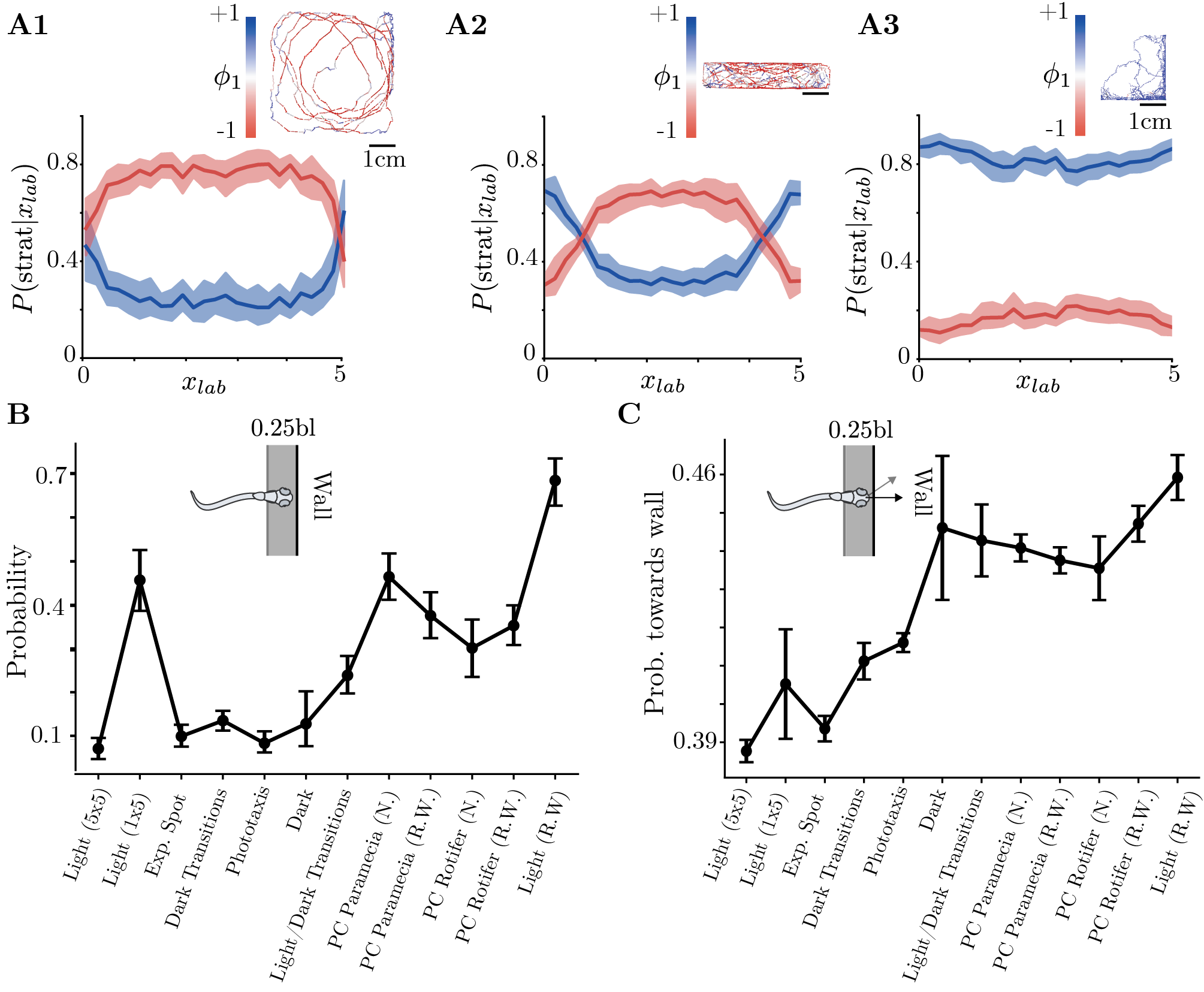}
\caption{\textbf{Spatial nature of the usage of Cruising-Wandering states, and details of the wall-following behavior across free-swimming conditions} \textbf{(A)} Probability of executing either Cruising or Wandering strategies conditioned on the position along the horizontal axis $x_\text{lab}$. As an inset, we plot an example trajectory of 500 bouts in each condition, color coded by $\phi_1$. \textbf{(A1)}  In a 5cm x 5cm arena, Wandering is enhanced only when the fish is at the corners of the arena, where it is forced to engage in a sequence of reorientations. \textbf{(A2)} In a 1cm x 5cm arena, similar to the 5cm x 5cm arena, the probability of Wandering increases only when the fish is at the corners. Shortening the $y-$axis of the arena prevents the fish from cruising along the vertical axis, whereas in the 5cm x 5cm arena fish could still cruise even for small and large values of $x_\text{lab}$. 
\textbf{(a3)} In a 2.5cm x 2.5cm arena, fish that were raised with rotifers engage in Wandering behavior throughout the entire arena. Notice also how in this case fish barely explore the center of the arena, mostly interacting with the walls while Wandering.
\textbf{(B)} Probability of thigmotaxis in each condition (here defined as the proportion of bouts in which the head position is on average smaller that a quarter of a body length away from the wall). Fish that were raised with Rotifers but are freely swimming have the highest probability of thigmotaxis, followed by the remaining prey capture conditions. In addition, we notice that the Light (1x5) condition the fish also spend time close to the wall, but they do so mostly by cruising along the major axis of the arena, see panel A2. 
\textbf{(C)} Probability of the reorienting towards the wall when the fish is close to the wall. Unlike the other Light conditions, fish raised with prey have a much higher probability of orienting towards the wall when close to it, indicating a different role to their thigmotatic behavior. Additionally, we find that in other prey capture conditions, as well as in the Dark and Ligh/Dark Tx. condition (in which Wandering strategies are also prevalent), fish also reorient towards the wall when they are close to it.}
\label{Suppl:6}
\end{figure}

\begin{figure}[h!]
\centering
\includegraphics[width = 1.\textwidth]{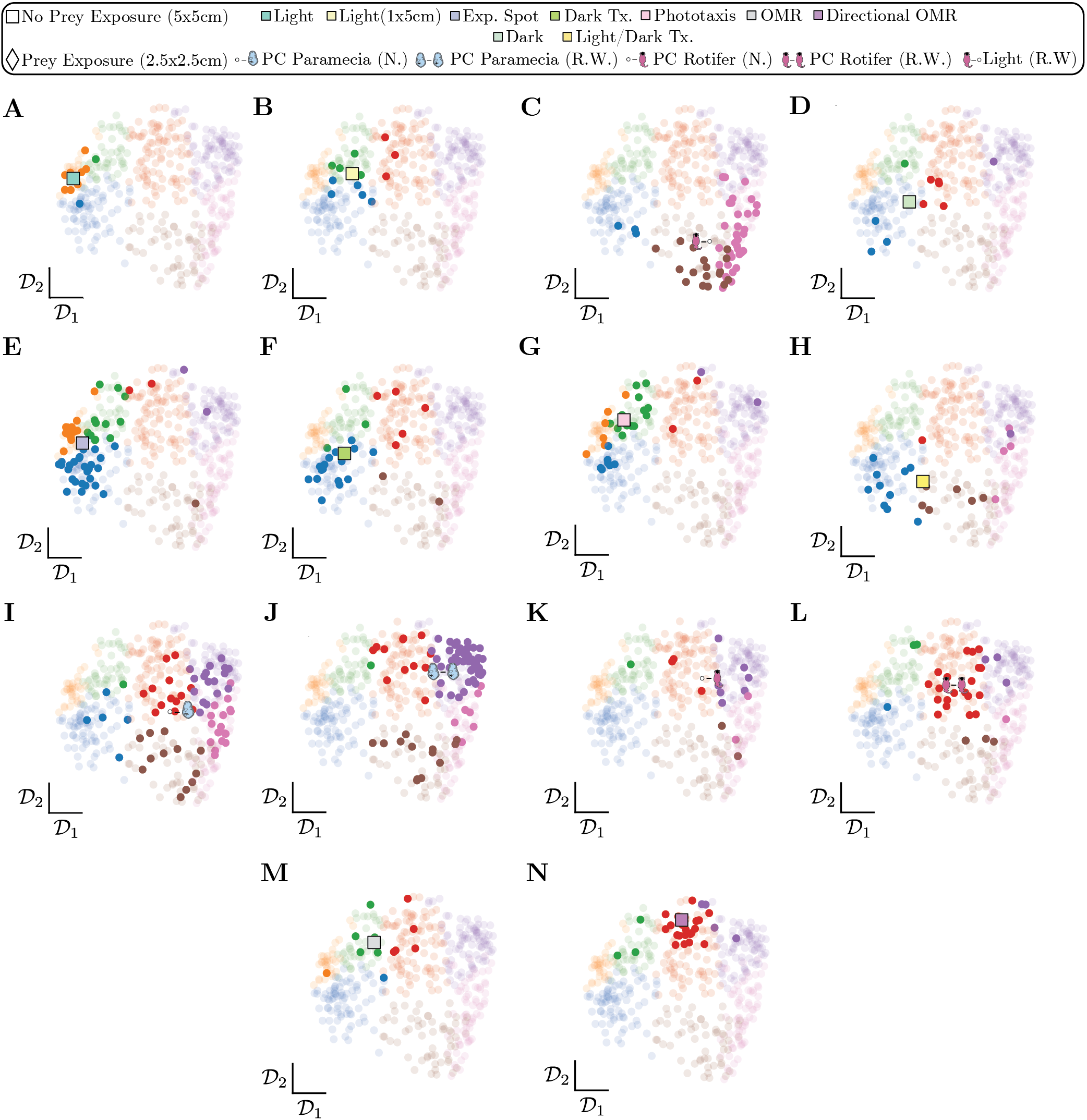}
\caption{\textbf{Variability among fish from different sensory conditions.} In each panel, we plot the position of each individual fish along $\mathcal{D}_1$ and $\mathcal{D}_2$ of the transition matrix space as the background, color coding each fish by their respective behavioral group. In addition, we highlight in darker colors the fish corresponding to each sensory condition, as well as the transition matrix obtained from all fish in a given condition through a different tick mark. \textbf{(A)} Light (5cm x 5cm) condition. \textbf{(B)} Light  (1cm x 5cm) condition  \textbf{(C)} Light R.W (2.5cm x 2.5cm) condition. \textbf{(D)} Dark (5cm x 5cm) condition. \textbf{(E)} Expanding Spot (5cm x 5cm) condition. \textbf{(F)} Dark transitions (5cm x 5cm) condition. \textbf{(G)} Phototaxis (5cm x 5cm) condition. \textbf{(H)} Light/Dark transitions (5cm x 5cm) condition. \textbf{(I)} Prey Capture Paramecia Naive (2.5cm x 2.5cm) condition. \textbf{(J)} Prey Capture Paramecia Raised With (2.5cm x 2.5cm) condition. \textbf{(K)} Prey Capture Rotifer Naive (2.5cm x 2.5cm) condition. \textbf{(L)} Prey Capture Paramecia Rotifer Raised With (2.5cm x 2.5cm) condition.
 \textbf{(M)} OMR (1cm x 5cm) condition. \textbf{(N)} Directional OMR (5cm x 5cm) condition.} 
\label{Suppl:7}
\end{figure}

\begin{figure}[h!]
\centering
\includegraphics[width = 1.\textwidth]{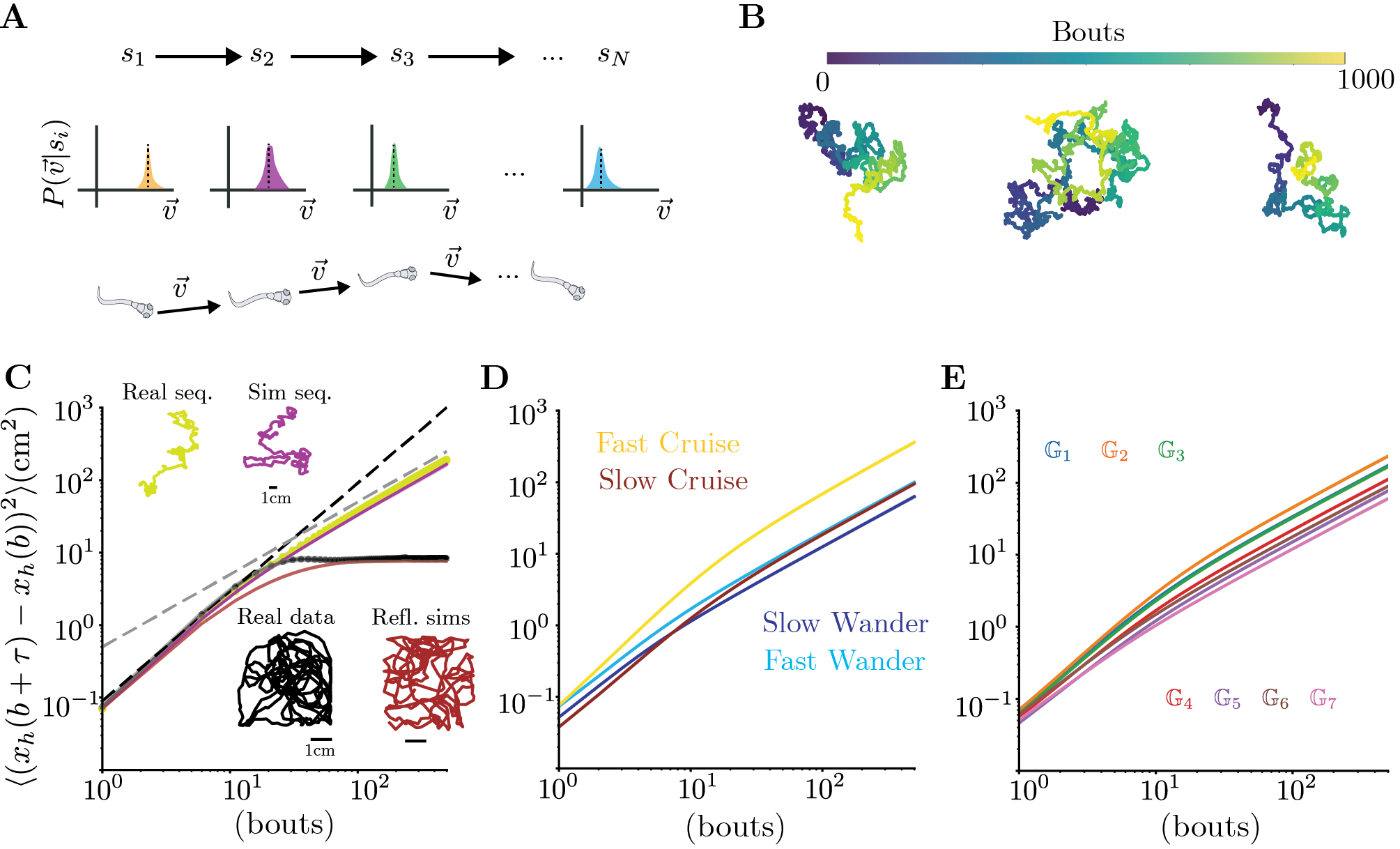}
\caption{\textbf{Assessing the statistical properties of the simulated trajectories through the mean squared displacement.} \textbf{(A)} Schematic of the simulation procedure. From a transition matrix among all $N^*=1200\,\text{states}$ and a discrete symbol $s_i$, we sample the next symbol $s_j$ from the $i$-th row of transition matrix, which corresponds to the conditional probability of observing any state $s_j$ given state $s_i$. For each symbol $s_i(t)$, we then sample a velocity vector from the distribution of velocities observed when fish where in state $s_i$ (see Methods for details). \textbf{(B)} Three example trajectories from the simulation procedure, each 1000 bouts long.
\textbf{(C)} We estimated the mean squared displacement (MSD) from the data (shown as a black scatter plot), as well as different kinds of trajectory simulations (colored lines). In the data, we observe super-diffusive behavior on short time scales, with the MSD scaling approximately as $\text{MSD} \sim \tau^{1.5}$ (black dashed line). However, the finite-size of the arenas quickly induces strong finite-size effects which bound the MSD from above. As a first assessment of the effects of the arena wall, we tried to reconstruct fish trajectories without boundary constraints by using the real symbolic sequence of each fish to generate new trajectories. We sample velocity vectors from the distribution of velocities obtained from the bouts corresponding to each symbol and use them to generate artificial trajectories, which we label as Real seq. (yellow), see schematic of Fig.\,\ref{Suppl:8}A and Methods for details. Notably, such trajectories match the statistics of the real data up to $\approx 20\,\text{bouts}$, at which point the data reaches its upper bound while the Real seq. trajectories start entering a diffusive regime in which $\text{MSD} \sim \tau$ (gray dashed line). We then generate trajectories from simulated symbolic sequences, which we call Sim seq. (purple). We find that the trajectories obtained from simulated symbolic sequences closely match the data on short timescales, but also match the trajectories obtained from the real symbolic sequence well across scales. Overall, we find that the arena sizes severely challenges an accurate estimate of the diffusive properties of larval zebrafish navigation, but that nonetheless our simulations match the statistics observed in the data within its limits. Note also that simply enforcing reflective boundary conditions (red) is not enough to capture the statistics, even at short timescales, pointing to an active interaction of the fish with the wall.
\textbf{(D)} Mean squared displacements of simulations from $N=4$ metastable strategies.  \textbf{(E)} Mean squared displacements of simulations from behavioral groups $\mathbb{G}_i, i\in[1,7]$.} 
\label{Suppl:8}
\end{figure}

\clearpage
\section*{References}
\bibliographystyle{bxv_abbrvnat}
\bibliography{refs.bib}
